%% file: CDMSII_LLH_Paper.tex
\documentclass[aps,reprint,showpacs,nofootinbib,superscriptaddress,floatfix]{revtex4-1}
\usepackage{amsmath,latexsym,amssymb,hyperref,graphicx}
\usepackage{times}
\usepackage{epsfig}
\usepackage{enumitem}
\usepackage{slashed}
\RequirePackage{lineno}
\hypersetup{colorlinks,citecolor= nicegreen,linkcolor= nicered}
\usepackage{color}
\definecolor{nicered}{rgb}{0.7,0.1,0.1}
\definecolor{nicegreen}{rgb}{0.1,0.5,0.1}

\newcommand{\be}{\begin{equation}}
\newcommand{\ee}{\end{equation}}
\newcommand{\bea}{\begin{eqnarray}}
\newcommand{\eea}{\end{eqnarray}}
\def\<{\langle}
\def\>{\rangle}

\begin{document}

\title[ Maximum Likelihood Analysis of Low Energy CDMS II Germanium Data ]{ Maximum Likelihood Analysis of Low Energy CDMS II Germanium Data  }

\affiliation{Division of Physics, Mathematics, \& Astronomy, California Institute of Technology, Pasadena, CA 91125, USA}
\affiliation{Fermi National Accelerator Laboratory, Batavia, IL 60510, USA}
\affiliation{Lawrence Berkeley National Laboratory, Berkeley, CA 94720, USA}
\affiliation{Department of Physics, Massachusetts Institute of Technology, Cambridge, MA 02139, USA}
\affiliation{Pacific Northwest National Laboratory, Richland, WA 99352, USA}
\affiliation{Department of Physics \& Astronomy, University of British Columbia, Vancouver BC, Canada  V6T 1Z1}
\affiliation{Department of Physics, Queen's University, Kingston ON, Canada K7L 3N6}
\affiliation{Department of Physics, Santa Clara University, Santa Clara, CA 95053, USA}
\affiliation{SLAC National Accelerator Laboratory/Kavli Institute for Particle Astrophysics and Cosmology, 2575 Sand Hill Road, Menlo Park 94025, CA}
\affiliation{Department of Physics, Southern Methodist University, Dallas, TX 75275, USA}
\affiliation{Department of Physics, Stanford University, Stanford, CA 94305, USA}
\affiliation{Department of Physics, Syracuse University, Syracuse, NY 13244, USA}
\affiliation{Department of Physics, Texas A\&M University, College Station, TX 77843, USA}
\affiliation{Departamento de F\'{\i}sica Te\'orica and Instituto de F\'{\i}sica Te\'orica UAM/CSIC, Universidad Aut\'onoma de Madrid, 28049 Madrid, Spain}
\affiliation{Department of Physics, University of California, Berkeley, CA 94720, USA}
\affiliation{Department of Physics, University of California, Santa Barbara, CA 93106, USA}
\affiliation{Department of Physics, University of Colorado Denver, Denver, CO 80217, USA}
\affiliation{Department of Physics, University of Evansville, Evansville, IN 47722, USA}
\affiliation{Department of Physics, University of Florida, Gainesville, FL 32611, USA}
\affiliation{Department of Physics, University of Illinois at Urbana-Champaign, Urbana, IL 61801, USA}
\affiliation{School of Physics \& Astronomy, University of Minnesota, Minneapolis, MN 55455, USA}
\author{R.~Agnese}\affiliation{Department of Physics, University of Florida, Gainesville, FL 32611, USA}
\author{A.J.~Anderson} \affiliation{Department of Physics, Massachusetts Institute of Technology, Cambridge, MA 02139, USA}
\author{D.~Balakishiyeva} \affiliation{Department of Physics, University of Florida, Gainesville, FL 32611, USA}
\author{R.~Basu~Thakur~} \affiliation{Fermi National Accelerator Laboratory, Batavia, IL 60510, USA}\affiliation{Department of Physics, University of Illinois at Urbana-Champaign, Urbana, IL 61801, USA}
\author{D.A.~Bauer} \affiliation{Fermi National Accelerator Laboratory, Batavia, IL 60510, USA}
\author{J.~Billard} \affiliation{Department of Physics, Massachusetts Institute of Technology, Cambridge, MA 02139, USA}
\author{A.~Borgland} \affiliation{SLAC National Accelerator Laboratory/Kavli Institute for Particle Astrophysics and Cosmology, 2575 Sand Hill Road, Menlo Park 94025, CA}
\author{M.A.~Bowles} \affiliation{Department of Physics, Syracuse University, Syracuse, NY 13244, USA}
\author{D.~Brandt} \affiliation{SLAC National Accelerator Laboratory/Kavli Institute for Particle Astrophysics and Cosmology, 2575 Sand Hill Road, Menlo Park 94025, CA}
\author{P.L.~Brink} \affiliation{SLAC National Accelerator Laboratory/Kavli Institute for Particle Astrophysics and Cosmology, 2575 Sand Hill Road, Menlo Park 94025, CA}
\author{R.~Bunker} \affiliation{Department of Physics, Syracuse University, Syracuse, NY 13244, USA}
\author{B.~Cabrera} \affiliation{Department of Physics, Stanford University, Stanford, CA 94305, USA}
\author{D.O.~Caldwell} \affiliation{Department of Physics, University of California, Santa Barbara, CA 93106, USA}
\author{D.G.~Cerdeno} \affiliation{Departamento de F\'{\i}sica Te\'orica and Instituto de F\'{\i}sica Te\'orica UAM/CSIC, Universidad Aut\'onoma de Madrid, 28049 Madrid, Spain}
\author{H.~Chagani} \affiliation{School of Physics \& Astronomy, University of Minnesota, Minneapolis, MN 55455, USA}
\author{Y.~Chen} \affiliation{Department of Physics, Syracuse University, Syracuse, NY 13244, USA}
\author{J.~Cooley} \affiliation{Department of Physics, Southern Methodist University, Dallas, TX 75275, USA}
\author{B.~Cornell} \affiliation{Division of Physics, Mathematics, \& Astronomy, California Institute of Technology, Pasadena, CA 91125, USA}
\author{C.H.~Crewdson} \affiliation{Department of Physics, Queen's University, Kingston ON, Canada K7L 3N6}
\author{P.~Cushman} \affiliation{School of Physics \& Astronomy, University of Minnesota, Minneapolis, MN 55455, USA}
\author{M.~Daal} \affiliation{Department of Physics, University of California, Berkeley, CA 94720, USA}
\author{P.C.F.~Di~Stefano} \affiliation{Department of Physics, Queen's University, Kingston ON, Canada K7L 3N6}
\author{T.~Doughty} \affiliation{Department of Physics, University of California, Berkeley, CA 94720, USA}
\author{L.~Esteban} \affiliation{Departamento de F\'{\i}sica Te\'orica and Instituto de F\'{\i}sica Te\'orica UAM/CSIC, Universidad Aut\'onoma de Madrid, 28049 Madrid, Spain}
\author{S.~Fallows} \affiliation{School of Physics \& Astronomy, University of Minnesota, Minneapolis, MN 55455, USA}
\author{E.~Figueroa-Feliciano} \affiliation{Department of Physics, Massachusetts Institute of Technology, Cambridge, MA 02139, USA}
\author{M.~Fritts} \affiliation{School of Physics \& Astronomy, University of Minnesota, Minneapolis, MN 55455, USA}
\author{G.L.~Godfrey} \affiliation{SLAC National Accelerator Laboratory/Kavli Institute for Particle Astrophysics and Cosmology, 2575 Sand Hill Road, Menlo Park 94025, CA}
\author{S.R.~Golwala} \affiliation{Division of Physics, Mathematics, \& Astronomy, California Institute of Technology, Pasadena, CA 91125, USA}
\author{M.~Graham} \affiliation{SLAC National Accelerator Laboratory/Kavli Institute for Particle Astrophysics and Cosmology, 2575 Sand Hill Road, Menlo Park 94025, CA}
\author{J.~Hall} \affiliation{Pacific Northwest National Laboratory, Richland, WA 99352, USA}
\author{H.R.~Harris} \affiliation{Department of Physics, Texas A\&M University, College Station, TX 77843, USA}
\author{S.A.~Hertel} \affiliation{Department of Physics, Massachusetts Institute of Technology, Cambridge, MA 02139, USA}
\author{T.~Hofer} \affiliation{School of Physics \& Astronomy, University of Minnesota, Minneapolis, MN 55455, USA}
\author{D.~Holmgren} \affiliation{Fermi National Accelerator Laboratory, Batavia, IL 60510, USA}
\author{L.~Hsu} \affiliation{Fermi National Accelerator Laboratory, Batavia, IL 60510, USA}
\author{M.E.~Huber} \affiliation{Department of Physics, University of Colorado Denver, Denver, CO 80217, USA}
\author{A.~Jastram} \affiliation{Department of Physics, Texas A\&M University, College Station, TX 77843, USA}
\author{O.~Kamaev} \affiliation{Department of Physics, Queen's University, Kingston ON, Canada K7L 3N6}
\author{B.~Kara} \affiliation{Department of Physics, Southern Methodist University, Dallas, TX 75275, USA}
\author{M.H.~Kelsey} \affiliation{SLAC National Accelerator Laboratory/Kavli Institute for Particle Astrophysics and Cosmology, 2575 Sand Hill Road, Menlo Park 94025, CA}
\author{A.~Kennedy} \affiliation{School of Physics \& Astronomy, University of Minnesota, Minneapolis, MN 55455, USA}
\author{M.~Kiveni} \affiliation{Department of Physics, Syracuse University, Syracuse, NY 13244, USA}
\author{K.~Koch} \affiliation{School of Physics \& Astronomy, University of Minnesota, Minneapolis, MN 55455, USA}
\author{A.~Leder} \affiliation{Department of Physics, Massachusetts Institute of Technology, Cambridge, MA 02139, USA}
\author{B.~Loer} \affiliation{Fermi National Accelerator Laboratory, Batavia, IL 60510, USA}
\author{E.~Lopez~Asamar} \affiliation{Departamento de F\'{\i}sica Te\'orica and Instituto de F\'{\i}sica Te\'orica UAM/CSIC, Universidad Aut\'onoma de Madrid, 28049 Madrid, Spain}
\author{R.~Mahapatra} \affiliation{Department of Physics, Texas A\&M University, College Station, TX 77843, USA}
\author{V.~Mandic} \affiliation{School of Physics \& Astronomy, University of Minnesota, Minneapolis, MN 55455, USA}
\author{C.~Martinez} \affiliation{Department of Physics, Queen's University, Kingston ON, Canada K7L 3N6}
\author{K.A.~McCarthy} \affiliation{Department of Physics, Massachusetts Institute of Technology, Cambridge, MA 02139, USA}
\author{N.~Mirabolfathi} \affiliation{Department of Physics, Texas A\&M University, College Station, TX 77843, USA}
\author{R.A.~Moffatt} \affiliation{Department of Physics, Stanford University, Stanford, CA 94305, USA}
\author{D.C.~Moore} \affiliation{Division of Physics, Mathematics, \& Astronomy, California Institute of Technology, Pasadena, CA 91125, USA}
\author{R.H.~Nelson} \affiliation{Division of Physics, Mathematics, \& Astronomy, California Institute of Technology, Pasadena, CA 91125, USA}
\author{S.M.~Oser} \affiliation{Department of Physics \& Astronomy, University of British Columbia, Vancouver BC, Canada  V6T 1Z1}
\author{K.~Page} \affiliation{Department of Physics, Queen's University, Kingston ON, Canada K7L 3N6}
\author{W.A.~Page} \affiliation{Department of Physics \& Astronomy, University of British Columbia, Vancouver BC, Canada  V6T 1Z1}
\author{R.~Partridge} \affiliation{SLAC National Accelerator Laboratory/Kavli Institute for Particle Astrophysics and Cosmology, 2575 Sand Hill Road, Menlo Park 94025, CA}
\author{M.~Pepin} \affiliation{School of Physics \& Astronomy, University of Minnesota, Minneapolis, MN 55455, USA}
\author{A.~Phipps} \affiliation{Department of Physics, University of California, Berkeley, CA 94720, USA}
\author{K.~Prasad} \affiliation{Department of Physics, Texas A\&M University, College Station, TX 77843, USA}
\author{M.~Pyle} \affiliation{Department of Physics, University of California, Berkeley, CA 94720, USA}
\author{H.~Qiu} \affiliation{Department of Physics, Southern Methodist University, Dallas, TX 75275, USA}
\author{W.~Rau} \affiliation{Department of Physics, Queen's University, Kingston ON, Canada K7L 3N6}
\author{P.~Redl} \affiliation{Department of Physics, Stanford University, Stanford, CA 94305, USA}
\author{A.~Reisetter} \affiliation{Department of Physics, University of Evansville, Evansville, IN 47722, USA}
\author{Y.~Ricci} \affiliation{Department of Physics, Queen's University, Kingston ON, Canada K7L 3N6}
\author{H.~E.~Rogers} \affiliation{School of Physics \& Astronomy, University of Minnesota, Minneapolis, MN 55455, USA}
\author{T.~Saab} \affiliation{Department of Physics, University of Florida, Gainesville, FL 32611, USA}
\author{B.~Sadoulet} \affiliation{Department of Physics, University of California, Berkeley, CA 94720, USA}\affiliation{Lawrence Berkeley National Laboratory, Berkeley, CA 94720, USA}
\author{J.~Sander} \affiliation{Department of Physics, University of South Dakota, Vermillion, SD 57069, USA}
\author{K.~Schneck} \affiliation{SLAC National Accelerator Laboratory/Kavli Institute for Particle Astrophysics and Cosmology, 2575 Sand Hill Road, Menlo Park 94025, CA}
\author{R.W.~Schnee} \affiliation{Department of Physics, Syracuse University, Syracuse, NY 13244, USA}
\author{S.~Scorza} \affiliation{Department of Physics, Southern Methodist University, Dallas, TX 75275, USA}
\author{B.~Serfass} \affiliation{Department of Physics, University of California, Berkeley, CA 94720, USA}
\author{B.~Shank} \affiliation{Department of Physics, Stanford University, Stanford, CA 94305, USA}
\author{D.~Speller} \affiliation{Department of Physics, University of California, Berkeley, CA 94720, USA}
\author{S.~Upadhyayula} \affiliation{Department of Physics, Texas A\&M University, College Station, TX 77843, USA}
\author{A.N.~Villano} \affiliation{School of Physics \& Astronomy, University of Minnesota, Minneapolis, MN 55455, USA}
\author{B.~Welliver} \affiliation{Department of Physics, University of Florida, Gainesville, FL 32611, USA}
\author{D.H.~Wright} \affiliation{SLAC National Accelerator Laboratory/Kavli Institute for Particle Astrophysics and Cosmology, 2575 Sand Hill Road, Menlo Park 94025, CA}
\author{S.~Yellin} \affiliation{Department of Physics, Stanford University, Stanford, CA 94305, USA}
\author{J.J.~Yen} \affiliation{Department of Physics, Stanford University, Stanford, CA 94305, USA}
\author{B.A.~Young} \affiliation{Department of Physics, Santa Clara University, Santa Clara, CA 95053, USA}
\author{J.~Zhang} \affiliation{School of Physics \& Astronomy, University of Minnesota, Minneapolis, MN 55455, USA}


\collaboration{SuperCDMS Collaboration}

\noaffiliation

\date{\today}

\smallskip 



\begin{abstract} 
We report on the results of a search for a Weakly Interacting Massive Particle (WIMP) signal in low-energy
data of the Cryogenic Dark Matter Search (CDMS~II) experiment using
a maximum likelihood analysis.  A background model is constructed using GEANT4 to
simulate the surface-event background from $^{210}$Pb decay-chain events, while
using independent calibration data to model the gamma background.  Fitting this background model to the data results in no statistically significant WIMP component. 
In addition, we perform fits using an analytic \textit{ad hoc} background model proposed by Collar and Fields, 
who claimed to find a large excess of signal-like events in our data. We confirm the strong preference for a signal 
hypothesis in their analysis under these assumptions, but excesses are observed in both single- and multiple-scatter events, 
which implies the signal is not caused by WIMPs, but rather reflects the inadequacy of their background model.
\end{abstract}


\maketitle



\section{Introduction}

\input{Introduction.tex}

\label{intro}
\section{CDMS~II Detectors}
\label{detectors_CDMSII}
\input{Detectors.tex}

\section{The Background model} 
\label{bg}
\input{BackgroundModel.tex}

\section{Maximum Likelihood analysis}
\label{MLA}
\input{MLH.tex}

\section{Systematics}
\label{systematics}
\input{Systematics.tex}

\section{Results}
\label{results}
\input{Results.tex}

\section{Comparisons to the Collar-Fields Style Fits}
\label{comp}
\input{ComparisonToCF.tex}
\section{Conclusion} 
\input{Conclusion.tex} 
\section{Acknowledgements}
\input{Acknowledgements.tex}

\bibliographystyle{ieeetr} 
\bibliography{Bibliography.bib}

\end{document}

%% file: Introduction.tex
The existence of dark matter has been confirmed through astrophysical
observations, most recently from the Planck collaboration finding that 27\% of
the universe consists of cold dark matter~\cite{Planck}. Weakly Interacting
Massive Particles (WIMPs)~\cite{WIMPsRef} are a favored candidate
to explain the dark matter. These might interact with normal matter by
elastically scattering from nuclei, but the scattering rates and WIMP masses are
unknown. To detect the nuclear recoil signals caused by WIMP scatters in
terrestrial detectors, interactions with normal matter that might mimic such
signals must be eliminated, or at least accurately modeled.  
Great care has been taken in the Cryogenic Dark Matter Search  
experiment (CDMS~II) to reduce the number of neutrons that would give nuclear recoil 
signals; less than 1 neutron-induced nuclear recoil is expected in the full dataset.
The dominant backgrounds arise from residual radioactivity in the materials used 
to construct the structures around the detectors; the decay products are typically 
photons causing electron recoil events.  

\begin{figure}
\centering
\includegraphics[width=0.49\textwidth]{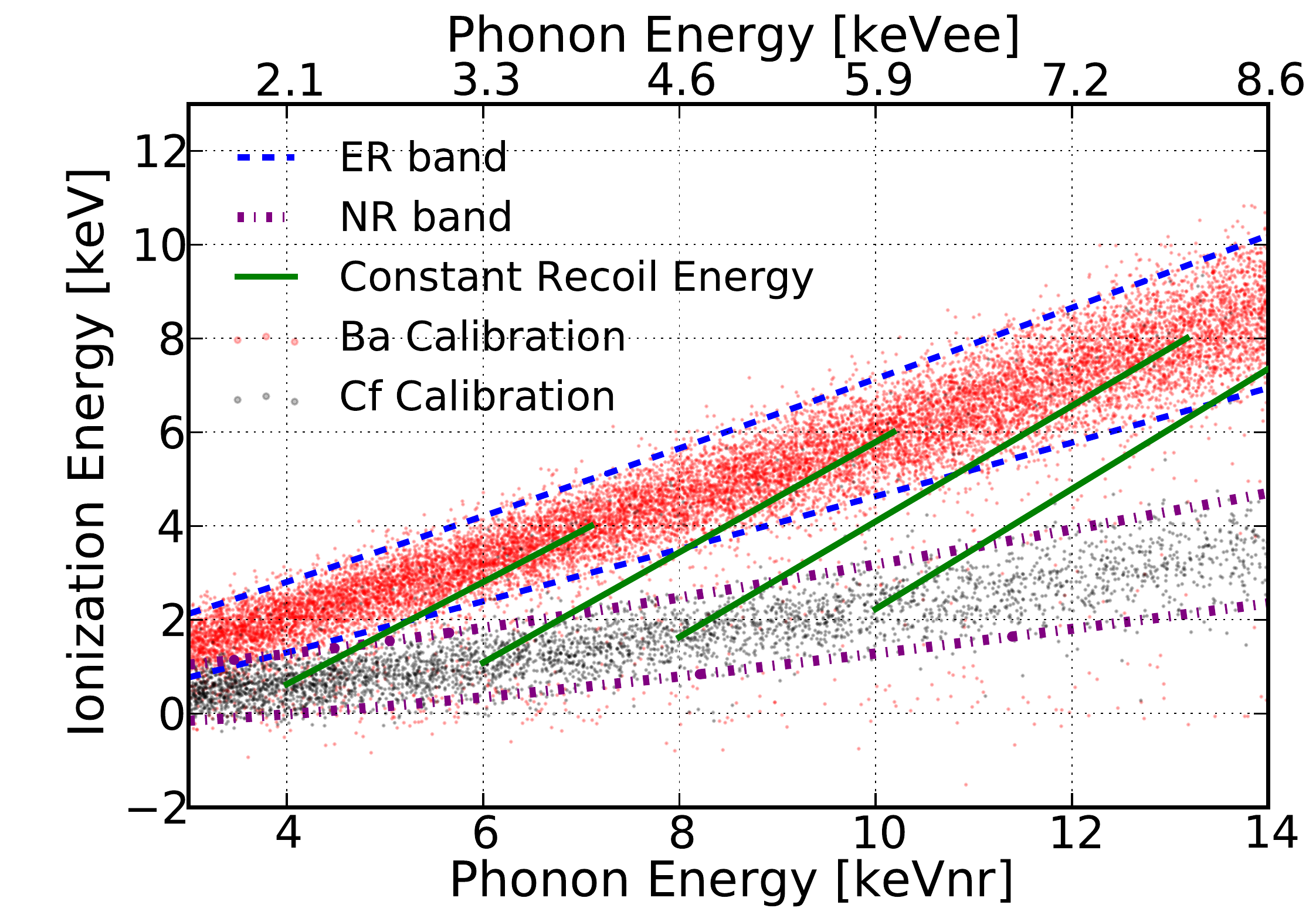}
\caption{(Color online)   Events in the ionization vs.~recoil-energy plane for 
one detector. Events from $^{133}$Ba and $^{252}$Cf calibration data are shown. The 
recoil energy scale is given by the total phonon energy minus the Neganov-Luke 
phonon \cite{Luke} contribution. The keVnr scale (bottom \textit{x} axis) gives 
the correct recoil 
energy at the center of the nuclear recoil (NR) band, while the keVee scale 
(top \textit{x} axis)  
gives the correct recoil energy at the center of the electron recoil (ER) band. }

\label{calibration_data}%
\end{figure}

CDMS~II~\cite{CDMSII_Soudan,CDMSII} cooled germanium and silicon detectors to
temperatures of $\lesssim$~50~mK in order to detect ionization and athermal 
phonons (`heat') generated by the elastic scattering of WIMPs from nuclei. 
Nuclear recoil (NR) events produce less ionization 
compared to similar energy electron recoil (ER) events. Consequently, NR and ER events 
can be separated.
However, at recoil energies $\lesssim$~10~keVnr (nuclear recoil equivalent energy, 
see Fig.~\ref{calibration_data}), background events start populating the signal 
region, as shown in
Fig.~\ref{calibration_data}. This figure shows calibration data for 
both $^{133}$Ba (a gamma source) and $^{252}$Cf (a neutron source).  
At energies above $\sim$10~keVnr 
there is good separation between gamma and neutron events. However, at lower 
energies the apparent bands of nuclear and electron recoils (NR and ER "bands") 
start to merge. 
At low energies a likelihood analysis can exploit 
the different distributions of signal and background in this two-dimensional space
to search for a WIMP signal.
We show that while this is a powerful technique, it requires a 
trustworthy background model. Producing such a  
model can be a challenging task.

Recent results from DAMA~\cite{DAMA}, CoGeNT~\cite{COGENT} and CDMS~II Si~\cite{CDMS_SI} 
can be interpreted as signals from 5--15~GeV/c$^2$ WIMPs, but results from 
CDMS~II Ge~\cite{modulation}, 
SuperCDMS~\cite{SuperCDMS_LT} and LUX~\cite{LUX} are in tension 
with these interpretations. 
Collar and Fields~\cite{CF} claimed evidence for a WIMP-like signal in CDMS~II data 
after attempting to estimate and effectively subtract the background.
We investigate that claim with a careful study of the backgrounds, 
thereby permitting an improved background modeling.
We present a maximum likelihood (ML) analysis of data from
CDMS~II's germanium detectors from 3--14~keVnr,
taken between 2006 and 2008. Details of the detectors are given 
in Section~\ref{detectors_CDMSII}.  We estimate  background distributions using
identified sources of background, either by simulating their detector response
or by using a representative calibration sample.  
A detailed description of the background model is provided in Section~\ref{bg} and 
its implementation in the ML analysis is discussed in Section~\ref{MLA}.
In Section~\ref{systematics} we study systematic effects before presenting the 
results in Section~\ref{results}.

In addition, we present a ML fit to these data using {\it ad hoc} analytic
models for background and signal.  We use the functional forms proposed by 
Collar and Fields~\cite{CF},  
whose fit to these data show a strong
preference (5.7$\sigma$ C.L.) for a model containing an exponentially falling
excess of events at low recoil energy in the NR band, consistent
with a low-mass WIMP hypothesis.  We also find a strong preference for a WIMP-like 
signal under the given assumptions. In  Section~\ref{comp} we show evidence that 
this is a consequence of the inadequacy of the background model.

For both types of ML analyses, we separately fit samples of events with energy 
deposited in only one detector (``single scatters" or ``singles") 
as well as samples of events with simultaneous energy depositions
in multiple detectors (``multiple scatters" or ``multiples"). 
A WIMP signal is not expected in the multiples data.  Therefore, a signal preference in the multiples data suggests that a similar excess in the singles data is likely caused by an incomplete or incorrect modeling of the backgrounds.

%% file: Detectors.tex
CDMS~II  used a mix of Ge and Si detectors, each $\sim$10~mm
thick and 76~mm in diameter.  They were packaged in copper housings that were 
stacked to form towers of six detectors each.  Here we chose to analyze the 
four Ge detectors, out
of 30 total (19~Ge and 11~Si), that had the most favorable electronic noise
characteristics as well as the lowest energy thresholds. 
The detectors chosen for this study are denoted T1Z1, T1Z5, T2Z5, and T3Z4, 
with ``T" indicating the tower number (1--5) and ``Z" indicating the detector 
number in the tower (1--6 from top to bottom).

\begin{figure}
\centering
\includegraphics[width=0.49\textwidth]{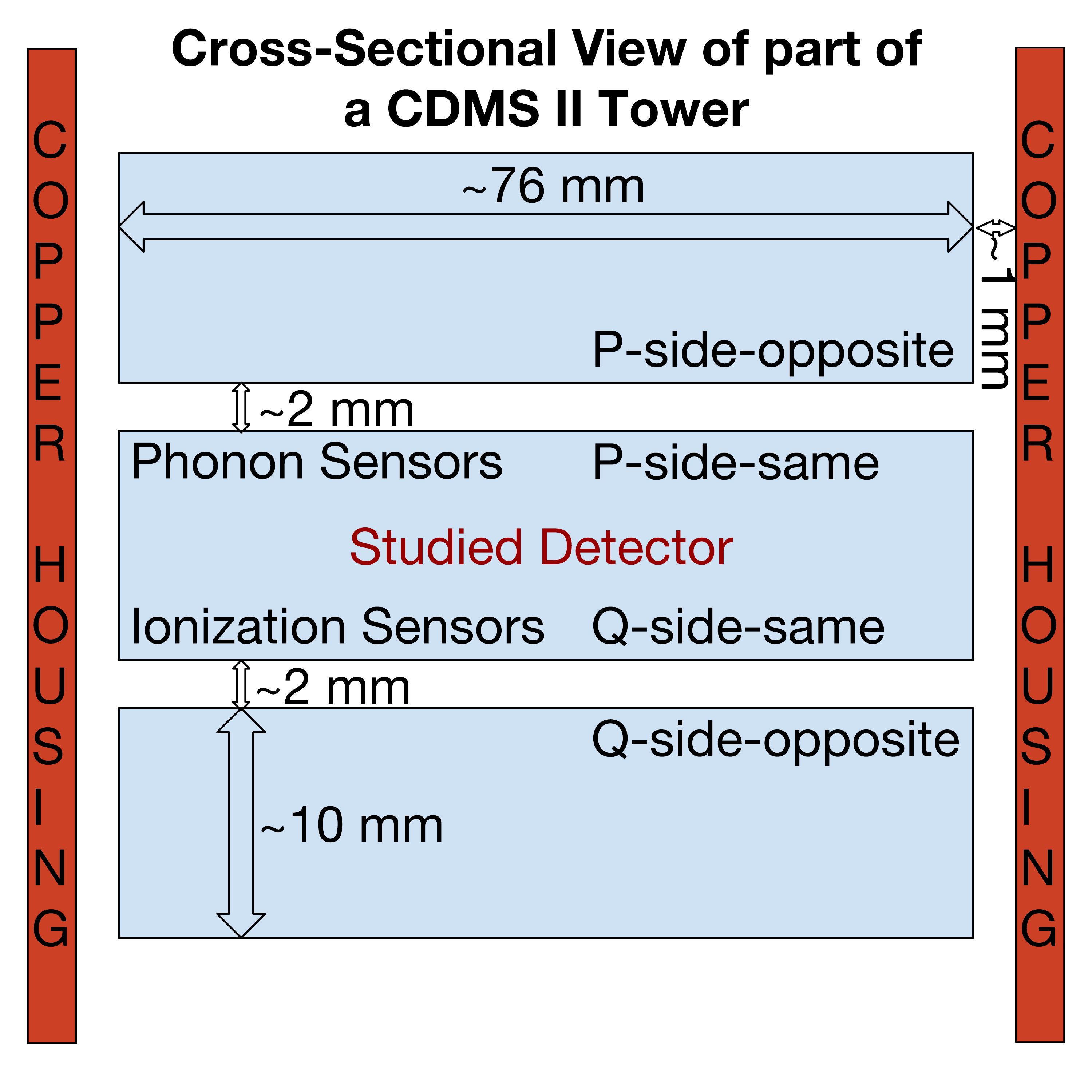}

\caption{A simple detector schematic (not to scale), showing an analyzed
detector in the center along with the two neighboring detectors. The detectors
are surrounded by copper housings. Detectors had phonon sensors on their top surfaces and ionization
sensors on the bottom. Surface events originating from decays on the 
``Studied Detector" are named ``P-side-same" or ``Q-side-same" depending on whether 
the decay occurred 
on the phonon or ionization side. In simulations, surface events on the 
``Studied Detector" that 
originate from a decay on a detector adjacent to the ``Studied Detector" are either named 
``P-side-opposite" or ``Q-side-opposite" depending on whether the decay caused an 
energy deposition on the 
phonon or ionization side of the Studied Detector.}
\label{schematicContribution}%
\end{figure}

The CDMS~II detectors were instrumented with phonon sensors on one surface and
ionization sensors on the other surface, while the sidewalls of the cylindrical
detectors were not instrumented. The ionization side had a central circular 
electrode and an outer guard ring that allowed differentiation of interactions
near the sidewall from those in the crystal interior.
A simple schematic of a detector is shown in Fig.~\ref{schematicContribution}.  
Simultaneous measurement of phonon and ionization signals enabled discrimination 
of NR and ER events via construction of the ratio of ionization to phonon 
energy (``ionization yield"). 
Events near a surface (depth~$<$~few~$\mu$m) can have diminished ionization yield, and thus ER events can leak into the NR signal region.
Events near the surfaces are
referred to as surface events, while events away from the surfaces are referred
to as bulk events.  More details of the CDMS~II detectors can be
found in~\cite{CDMSII_Soudan,CDMSII}.

%% file: BackgroundModel.tex
The CDMS~II detectors were shielded from external backgrounds with layers of
copper, lead and polyethylene.  Furthermore, to decrease the background from neutrons
produced by cosmic rays,  the experiment was located 2090~meters water 
equivalent underground at the Soudan Underground Laboratory and surrounded 
by a muon veto detector. However, the detectors are not background-free for the 
lowest-energy recoils considered here.  In particular, we consider two types of 
ER background:  the ``gamma background" and 
the ``surface-event background."  The former results from scatters of gamma 
rays throughout the detector and enters the low-energy signal region where 
finite energy resolution causes the ER and NR bands to overlap.  The 
surface-event background is due to events near the detector faces and sidewalls that 
have incomplete charge collection, resulting in degraded ionization yield that leads 
to misidentification as NR events. The neutron background is very low 
($<$~1 event in this dataset) and is therefore ignored.

\subsection{The Gamma Background}

\begin{figure}
\centering
\includegraphics[width=0.49\textwidth]{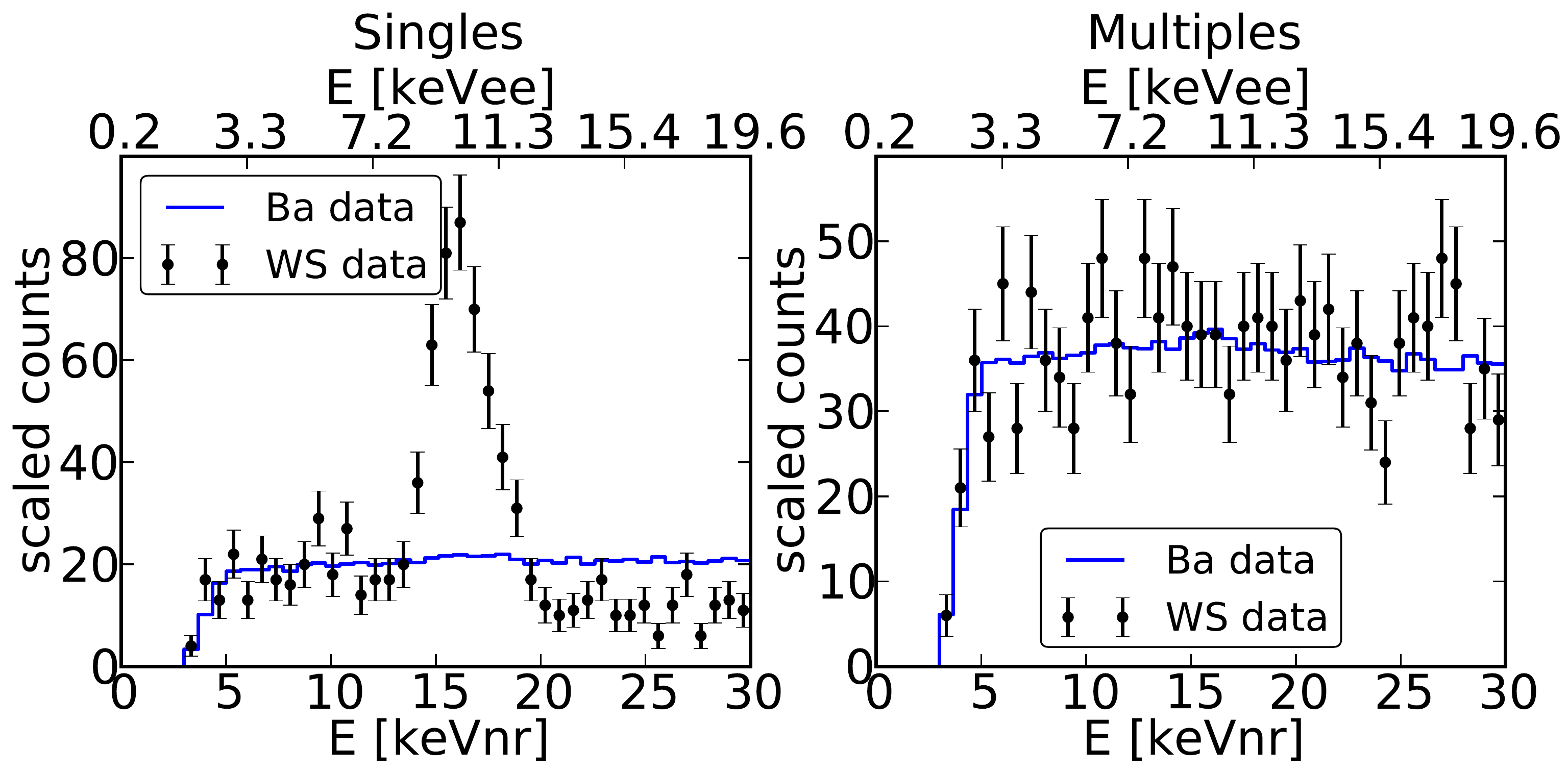}
\caption{(Color online) WIMP-search (WS) data and Ba calibration data for events 
within the ER band for detector T1Z2, given in NR (ER) equivalent energy along the 
bottom (top) axis. The 10.4 keVee (17 keVnr) Ge activation 
line is clearly seen in the single-scatter WIMP-search data (left panel) but absent in 
the Ba calibration and multiple-scatter WIMP-search data (right panel).}
\label{bavsdata_largeRange}%
\end{figure}

All materials contain radioactive contaminants.  Thus, although care was 
taken to minimize radiocontamination in the construction of the CDMS~II cryostat, 
support structures, detectors housings, and the detectors themselves, 
each component contains some contamination.
The majority of the gamma background observed in CDMS~II is caused by decays 
from radioactive U, Th (and their decay chain daughters) and $^{40}$K   
occurring in the surrounding materials. Additionally, Ge has radioactive isotopes that 
can be produced by neutrons or cosmogenic radiation ($^{68}$Ge and $^{71}$Ge). 
These isotopes decay via electron capture producing characteristic lines 
at 10.4~keV (K-shell) and 1.3~keV (L-shell). We chose our analysis energy 
range of 3--14~keVnr to avoid these activation lines.
Figure~\ref{bavsdata_largeRange} 
shows the gamma background for a recoil energy up to 30~keVnr for both single and 
multiple scatters, with the K-shell activation line clearly visible at 
$\sim$17~keVnr in the single-scatter data.
Other low-energy electron recoils (or ``gammas") result from a variety of 
sources, including cosmogenic activation of non-Ge isotopes and small-angle
Compton scattering.

\subsection{The Gamma Background Model}
\label{Gamma_background}
\begin{figure}
\centering
\includegraphics[width=0.49\textwidth]{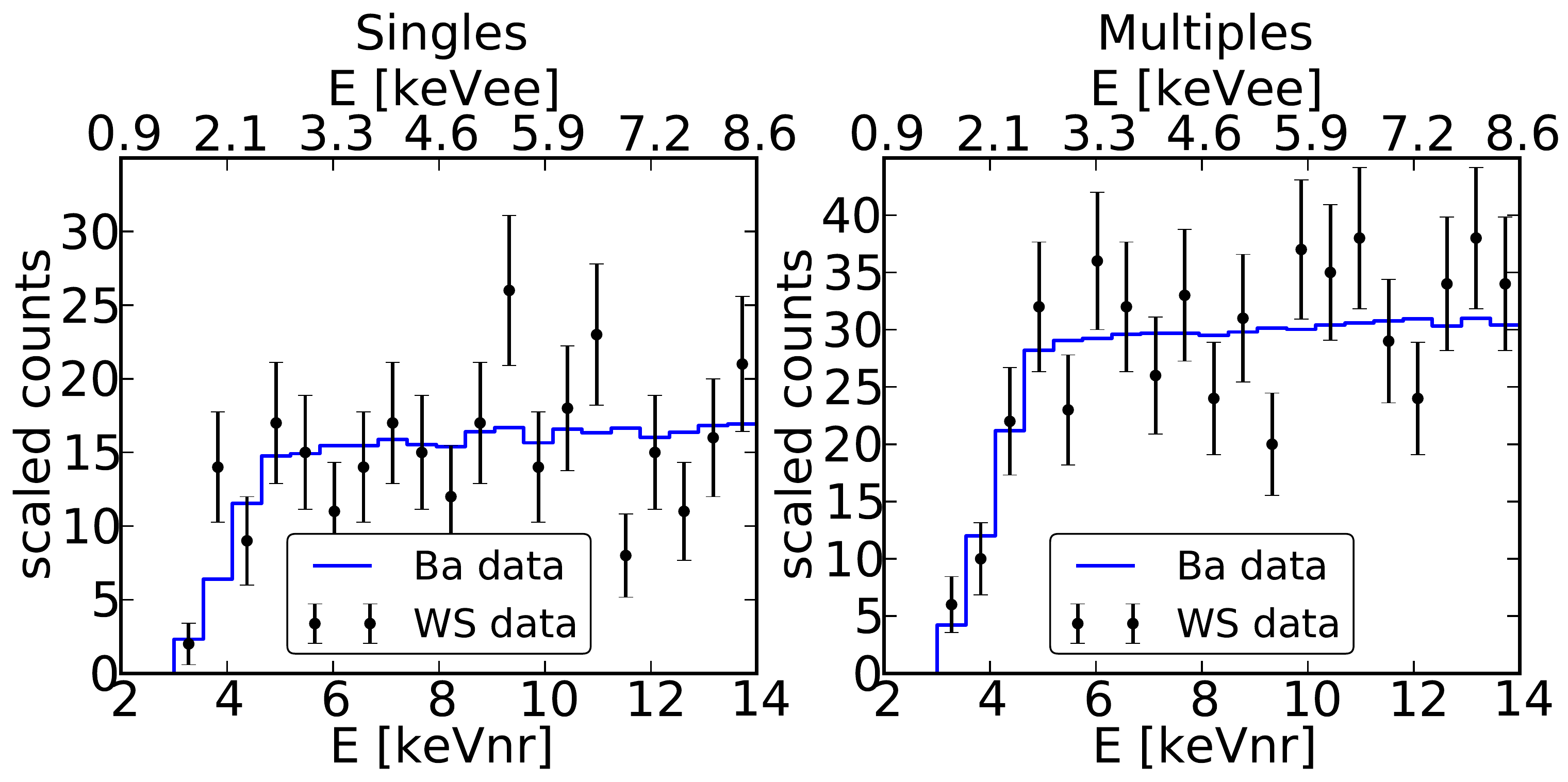}
\caption{(Color online)  Spectral comparison of events in the ER band. 
The singles spectrum is shown in the left panel while the
multiples spectrum is shown in the right panel. Ba calibration and WIMP-search
(WS) data are well matched. The p-value from a  Kolmogorov–-Smirnov test 
for the detector shown (T1Z2) is 0.8 (0.5) for multiples
(singles). The p-values for the other detectors are stated in the text.}
\label{bavsdata}%
\end{figure}

\begin{figure}
\centering
\includegraphics[width=0.49\textwidth]{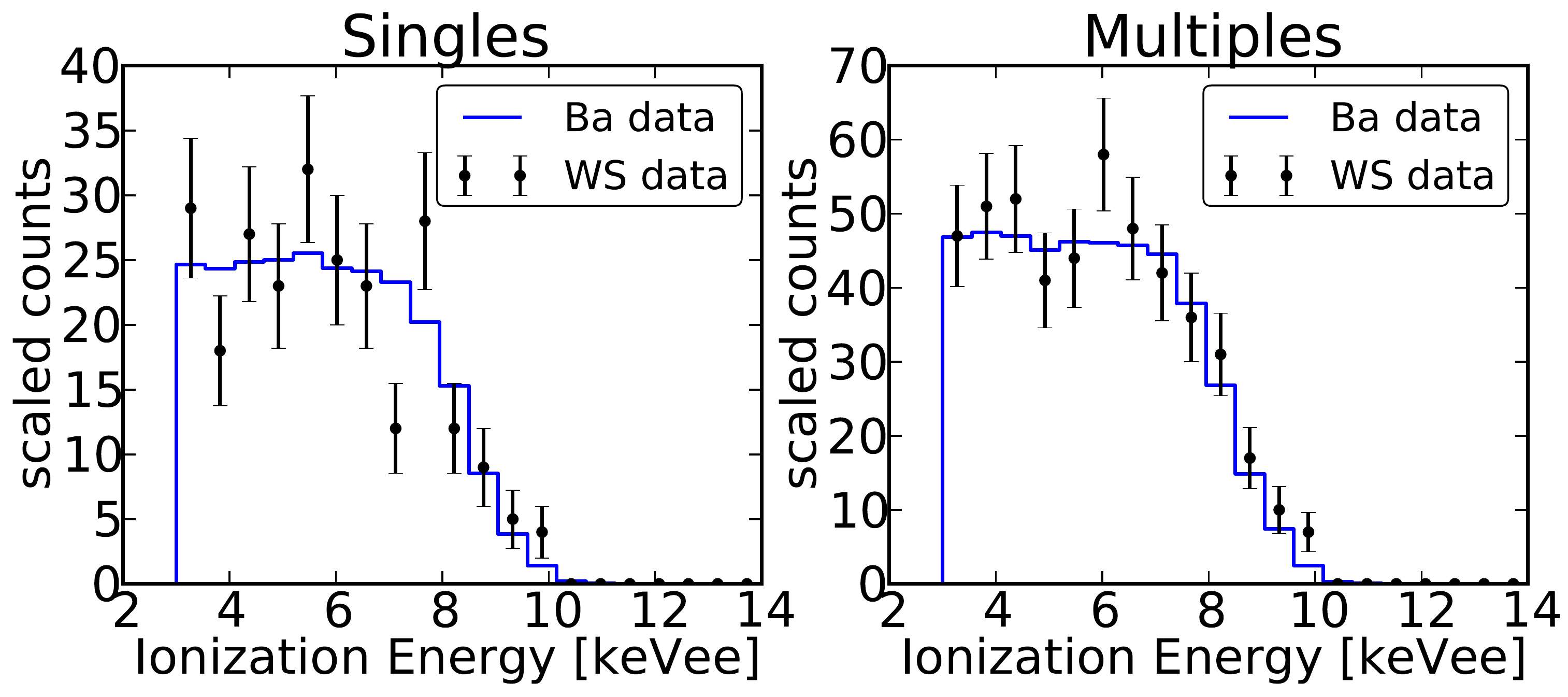}
\caption{(Color online) Comparison of the ionization-energy spectra for events 
in the ER band. The singles spectrum is shown in the left
panel while the multiples spectrum is shown in the right panel. Ba calibration
and WIMP-search (WS) data are well matched. The p-value from a Kolmogorov–-Smirnov test 
 for the detector shown (T1Z2) is 0.2 (0.6) for
multiples (singles). The p-values for the other detectors are stated in the text.}
\label{bavsdata_q}%
\end{figure}

Bulk gamma events are modeled using $^{133}$Ba calibration data.
Although dominated by a line at 356~keVee, sufficient Compton scattering 
occurs throughout the surrounding mechanical structures that a flat recoil spectrum
is observed between 3 and 14~keVnr (see Fig.~\ref{bavsdata}). In order for the Ba calibration 
data to be a good proxy for WIMP-search (all data taken without a calibration source) gamma events, the energy spectrum of WIMP-search  
and barium calibration events must be the same in the energy region of interest.
Figure~\ref{bavsdata} shows a comparison between the energy spectrum of barium
and WIMP-search events in the ER band. 
Kolmogorov--Smirnov (KS) tests \cite{KS}  comparing the two distributions between 
3 and 14~keVnr indicate that their shapes are compatible. The individual detectors 
T1Z2, T1Z5, T2Z5 and T3Z4 have KS p-values of 0.8 (0.5), 0.7 (0.2), 0.07 (0.8), 
and 0.013 (0.8) for multiples (singles), respectively.
Differences in ionization energy between Ba and WIMP-search data in the ER 
band may result in systematic effects.
Figure~\ref{bavsdata_q} compares the ionization-energy spectra inside the ER band.  Again, 
KS-test p-values indicate that the Ba and WIMP-search spectral shapes are 
compatible (for both singles and multiples). The 
individual detectors (in the same order) have KS p-values of 0.2 (0.6), 
0.02 (0.2), 0.7 (0.5), 0.8 (0.9) for multiples (singles), respectively. This provides assurance that 
any ionization-yield differences between the two data types have minimal 
influence on the modeling of the gamma background in the ER band.  

\subsection{The Surface-Event Background}
Surface events are defined as particle interactions that occur within a few
$\mu$m of the surface of a detector. Such events can have diminished charge 
collection and can even result in a complete loss of the ionization signal, 
in which case they are called zero-charge (ZC) events. 
The majority of surface events come from decays of the $^{210}$Pb decay chain, 
a long-lived 
product of the ubiquitous $^{222}$Rn whose daughters implant into 
surfaces during fabrication 
of the detectors and housings~\cite{Pb206_Redl}. 
The $^{210}$Pb decay chain produces relatively low-energy decay products
that do not penetrate the detectors deeply enough to have full charge
collection, leading to a significant number of surface or ZC events.
 
\subsection{The Surface-Event Model}
\label{SE}

\begin{figure}
\centering
\includegraphics[width=0.5\textwidth]{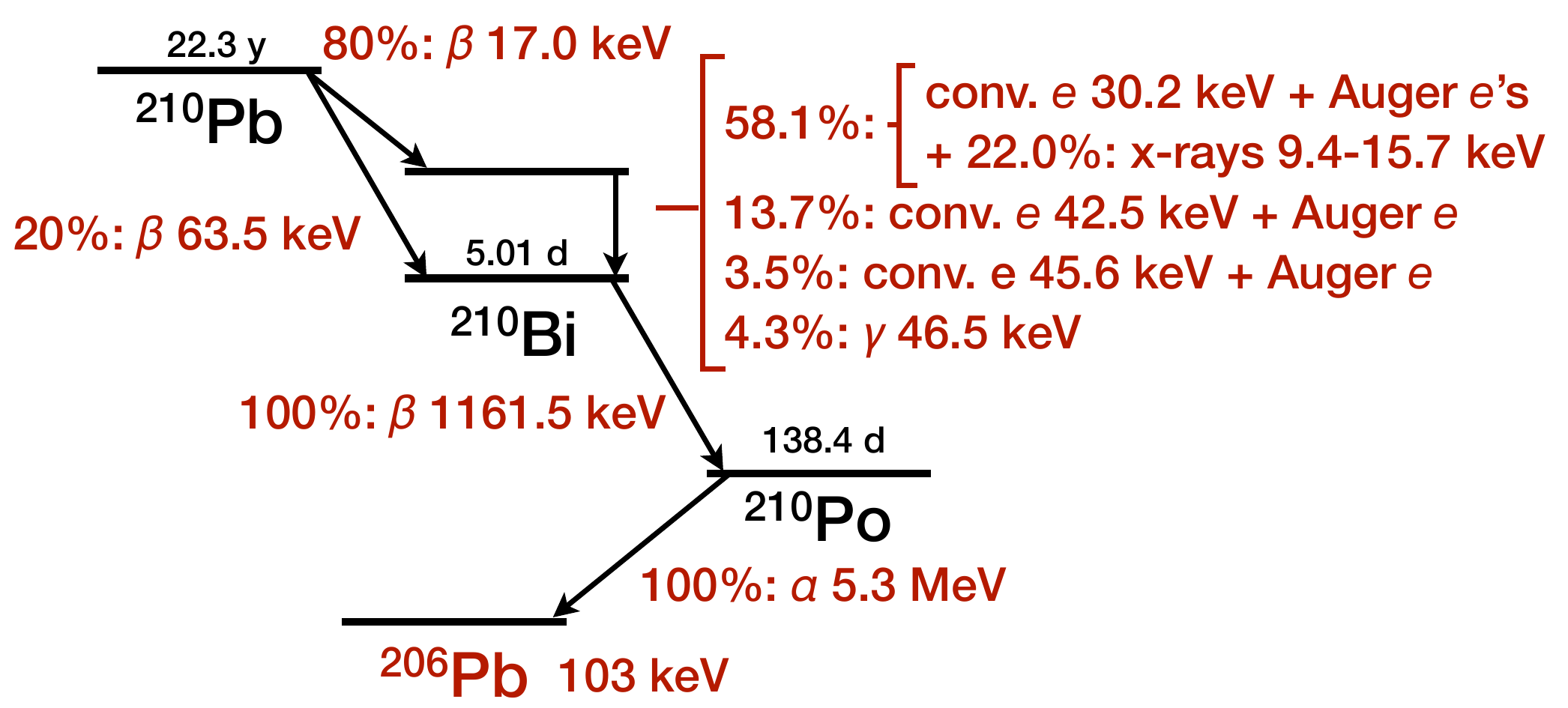}
\caption{(Color online) The dominant decay modes of the $^{210}$Pb decay chain. The 
alpha decay concluding this decay chain gives the $^{206}$Pb nucleus 103~keV 
of recoil energy.}
\label{decayChain}%
\end{figure}

We start our GEANT4~\cite{Geant-B} simulation of the surface-event background 
by contaminating the
surface of both the Cu detector housings and the Ge detectors with $^{214}$Po
nuclei that are allowed to decay isotropically. In addition to using the standard
GEANT4 physics lists it is imperative that the ``Screened Nuclear Recoil Physics List"
(SNRPL)~\cite{iZipRejection,SNR,Pb206_Redl} is invoked in order to correctly simulate
implantation of heavy, low-energy ($\lesssim$~500~keV) nuclei. The
SNRPL is based on algorithms used in SRIM~\cite{SRIM} and has been confirmed
to produce compatible results~\cite{SNR}. After the initial implantation of the
$^{210}$Pb nuclei, we simulate the full $^{210}$Pb decay chain shown in Fig.~\ref{decayChain}. 
The initial $^{210}$Pb decay produces a mix of electrons and low-energy photons, most of
which are sufficiently low in energy to be classified as surface events
\cite{iZipRejection,Pb206_Redl}. The $^{210}$Bi beta decay has an endpoint of
$\sim$1.2~MeV. In this analysis we focus on low-energy events (below
$\sim$14~keV) and therefore only a small fraction of $^{210}$Bi decays will
fall into our signal region, making the $^{210}$Bi a sub-dominant component. 
The final decay in the $^{210}$Pb decay chain is
another Po-isotope alpha decay; $^{210}$Po decays to $^{206}$Pb, producing a
5.3~MeV alpha particle. The alpha particle is unlikely to
contribute to our background because of its high energy.  
The 103~keV recoil energy of the $^{206}$Pb nucleus, however, may be degraded 
sufficiently to appear in the low-energy signal region because it may have to 
travel some distance through the surface in which the parent $^{210}$Po atom 
is implanted~\cite{iZipRejection,Pb206_Redl}. The number of particles produced
in this decay chain is proportional to the number of alpha particles produced in the
$^{210}$Po decay, so the observed $^{210}$Po alpha rate (measured 
from a high-energy sideband in the WIMP-search data) 
is used to estimate the total number of events expected to be observed from
$^{210}$Pb decay chain products (in the low-energy signal region) for each detector.

\begin{figure*}
\centering
\includegraphics[width=1.0\textwidth]{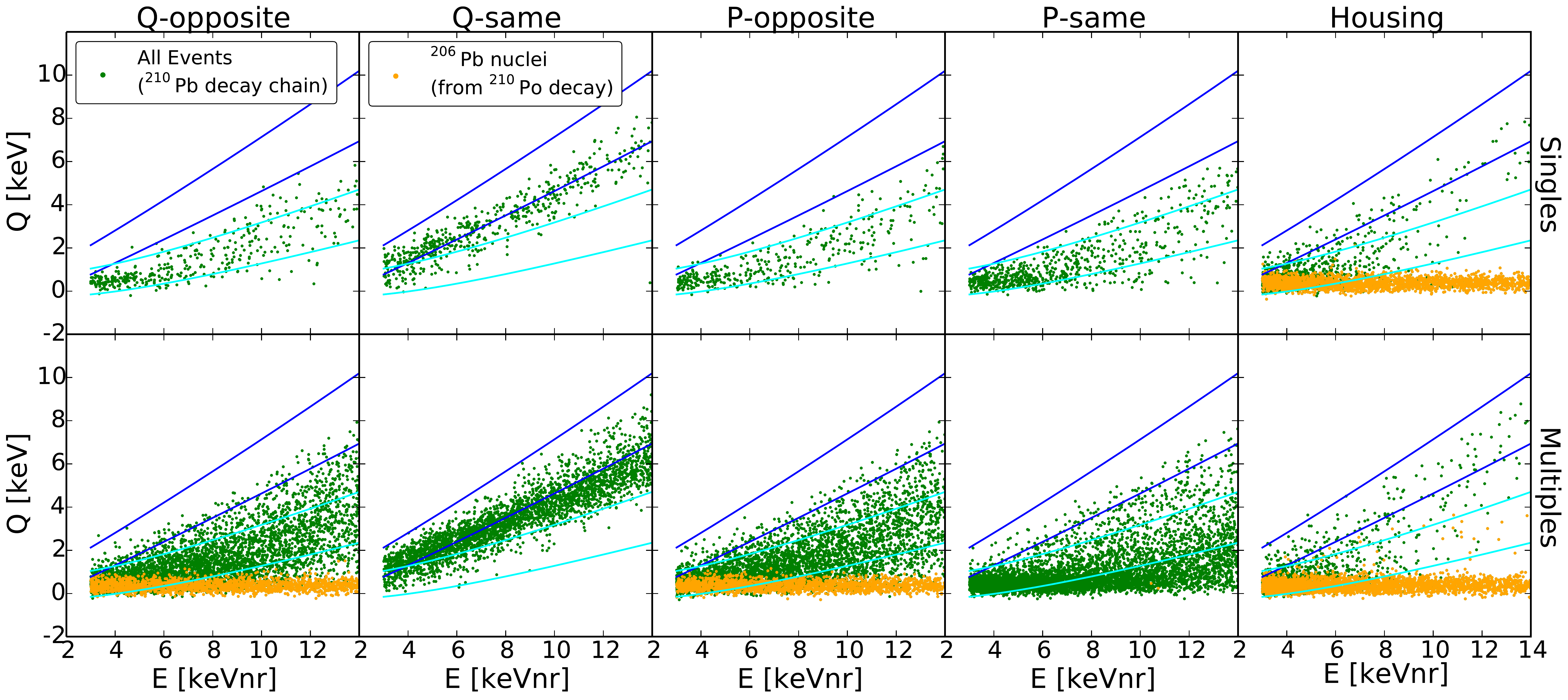}
\caption{(Color online, legend applies to all panels) Expected distribution of background
events in the plane of ionization versus~recoil energy (for a generic CDMS~II Ge detector), 
from simulations of surface events originating 
from the adjacent detector and striking the detector's charge side (far left column), 
from the detector's charge side (second column from left), from the adjacent 
detector and striking the detector's 
phonon side (third column), from the detector's phonon side (fourth column), 
and from the housing (last column), 
as labeled (see Fig.~\ref{schematicContribution} for a schematic). 
The top row shows single-scatter events and the bottom shows multiple-scatter
events. The upper dark (blue) pair of curves represents the electron recoil 
band while the nuclear 
recoil band is shown by the lower, lighter (teal) pair of curves. Events 
from $^{210}$Po decays that produce nuclear recoils are
highlighted in a lighter shade (orange) at Q near zero. 
This simulation has $\sim$100$\times$ more events than expected in the WIMP-search data.
Note that the relative numbers of events in each plot are fixed here; 
none of the relative normalizations (either of the 5 components or 
of singles to multiples) are allowed to float. Furthermore,
note that decays on different surfaces cause quite different spectra, 
which need to be considered 
for a reliable background model.}
\label{SimContribution}%
\end{figure*}

As described above, GEANT4~\cite{Geant-B} is utilized to 
simulate the particle interactions
in our detectors. However, the standard GEANT4 framework is currently not 
capable of simulating the phonons and electron-hole pairs produced 
by particle interactions in semiconductor crystals (i.e., the detailed 
detector response), and therefore this
estimation must be made after the GEANT4 simulation completes. We  
extended the GEANT4 framework to include these processes~\cite{Geant4_sim}, however
we did not use this new software here since it would have gone beyond the scope of this paper. 
The amount of phonon and ionization energy collected by the sensors is also not 
modeled with GEANT4 and
must be done post-simulation. Consider a particle interaction that produces an
initial combination of phonons and electron-hole pairs. The phonons diffuse
through the crystal, and the electron-hole pairs are drifted to the sensors
using a small drift field ($\sim$~3~V/cm), emitting additional phonons on the
way. The amount of charge collected depends on a few factors. The first 
(and most obvious) is the absolute number of electron-hole pairs produced by 
an event. For events producing recoiling electrons (from incident photons 
and electrons), one electron-hole pair escapes the interaction 
region per 3.0~eV of deposited recoil energy (on average)~\cite{eEV_cite}. 
Events recoiling
off nuclei produce fewer charge carriers, with the amount given approximately by 
standard Lindhard theory \cite{Lindhard,CDMSII}.
A particle hitting a
detector near a surface (within $\sim$~1~$\mu$m) will have suppressed charge
collection as well~\cite{DeadLayer}. In Section~\ref{systematics} we show evidence 
that for the detector sidewalls this depth scale is likely a factor of 5 smaller 
than for the detector faces. This led us to systematically vary the sidewall surface depth in the 
limit calculation presented in Section~\ref{results} to account for 
systematic uncertainties. For an event 
that occurs further than 
$\sim$~1~$\mu$m away
from a surface most of the produced charge is collected. For events on the side
instrumented for phonon readout (henceforth referred to as the ``phonon side'')
or on a sidewall we model the amount of charge collected to exponentially go to
zero at the surface, while for events on the side instrumented for ionization readout
(henceforth referred to as the ``ionization side'') we collect a
minimum of $\sim$~50$\%$ of the produced charge carriers, exponentially increasing
to 100$\%$ with a characteristic length of 1~$\mu$m (see Section~\ref{systematics}). 
Considering that the ionization and phonon sides of the detectors have different
charge collection characteristics, it is possible to separate the simulation into the
five components shown in Fig.~\ref{schematicContribution}; 
\textbf{1:} Events that originate in the Cu housings (Housing),
\textbf{2:} Events that originate on the detector currently being studied, on
either the charge side (Q-same) or \textbf{3:} the phonon side (P-same).
Events can also originate either from \textbf{4:} the detector adjacent
to the charge side (Q-opposite), or \textbf{5:} the detector adjacent to the
phonon side (P-opposite). Figure~\ref{SimContribution}  shows how 
these components contribute to the overall event distribution in the ionization- versus 
recoil-energy plane. 
To obtain a more realistic detector response, electronic noise 
(as measured with calibration data) was added to the simulated 
ionization and phonon energies.

\subsection{The Full Background Model}
\label{FullBackgroundModel}
\begin{figure}
\centering
\includegraphics[width=0.48\textwidth]{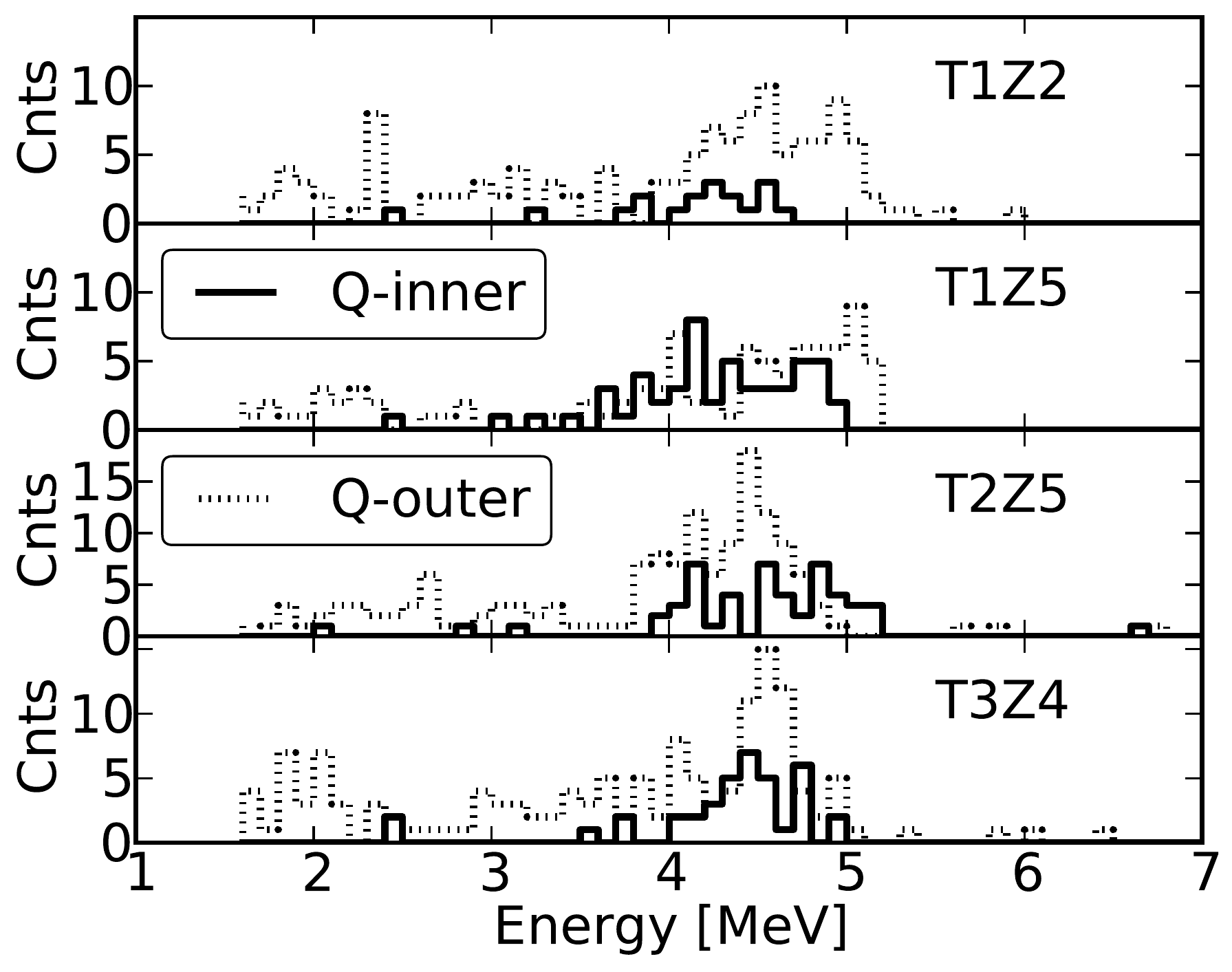}
\caption{Spectra of alpha events observed in (from top to bottom) detector 
T1Z2, T1Z5, T2Z5, and T3Z4.
Our analysis assumes that any alpha with
energy $>$~4~MeV is from a $^{210}$Po decay. 
Shown are alpha events detected on the inner (``Q-inner", solid) and
outer (``Q-outer", dashed) electrodes. Q-outer events most likely
originate from the copper housings. }
\label{AlphaEventPlot}%
\end{figure}%
\begin{figure*}
\centering
\includegraphics[width=0.49\textwidth]{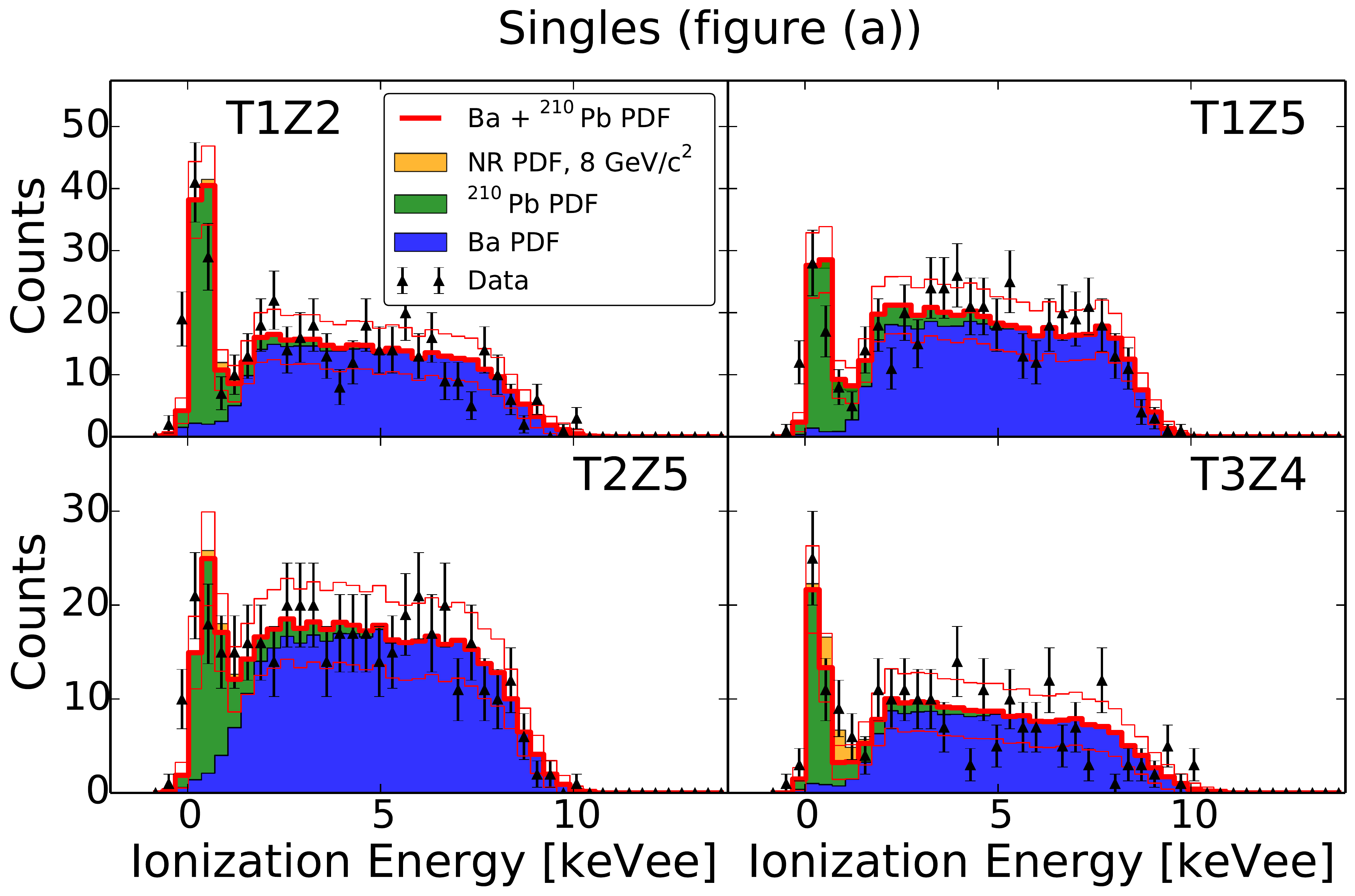}
\includegraphics[width=0.49\textwidth]{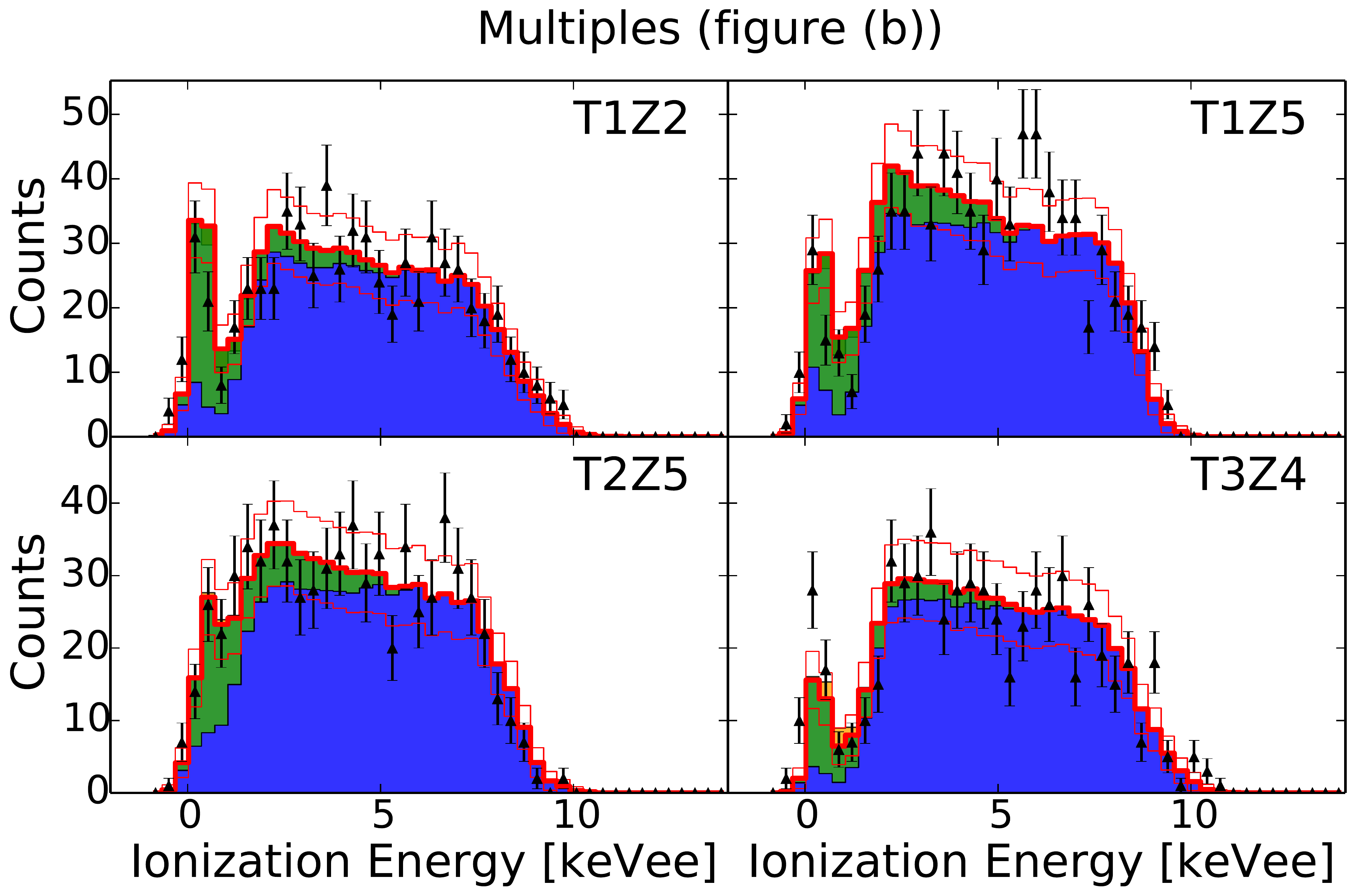}
\includegraphics[width=0.49\textwidth]{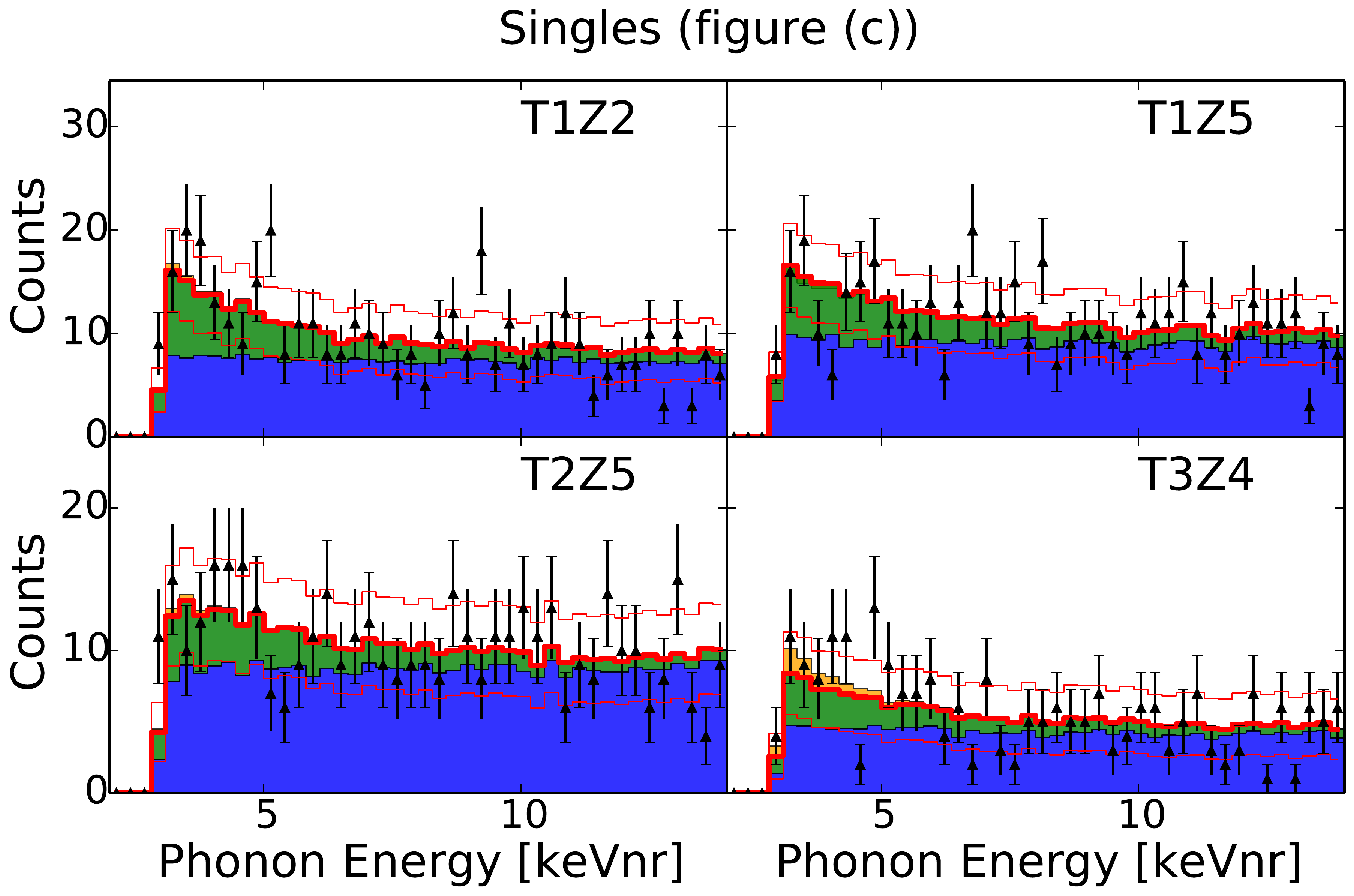}
\includegraphics[width=0.49\textwidth]{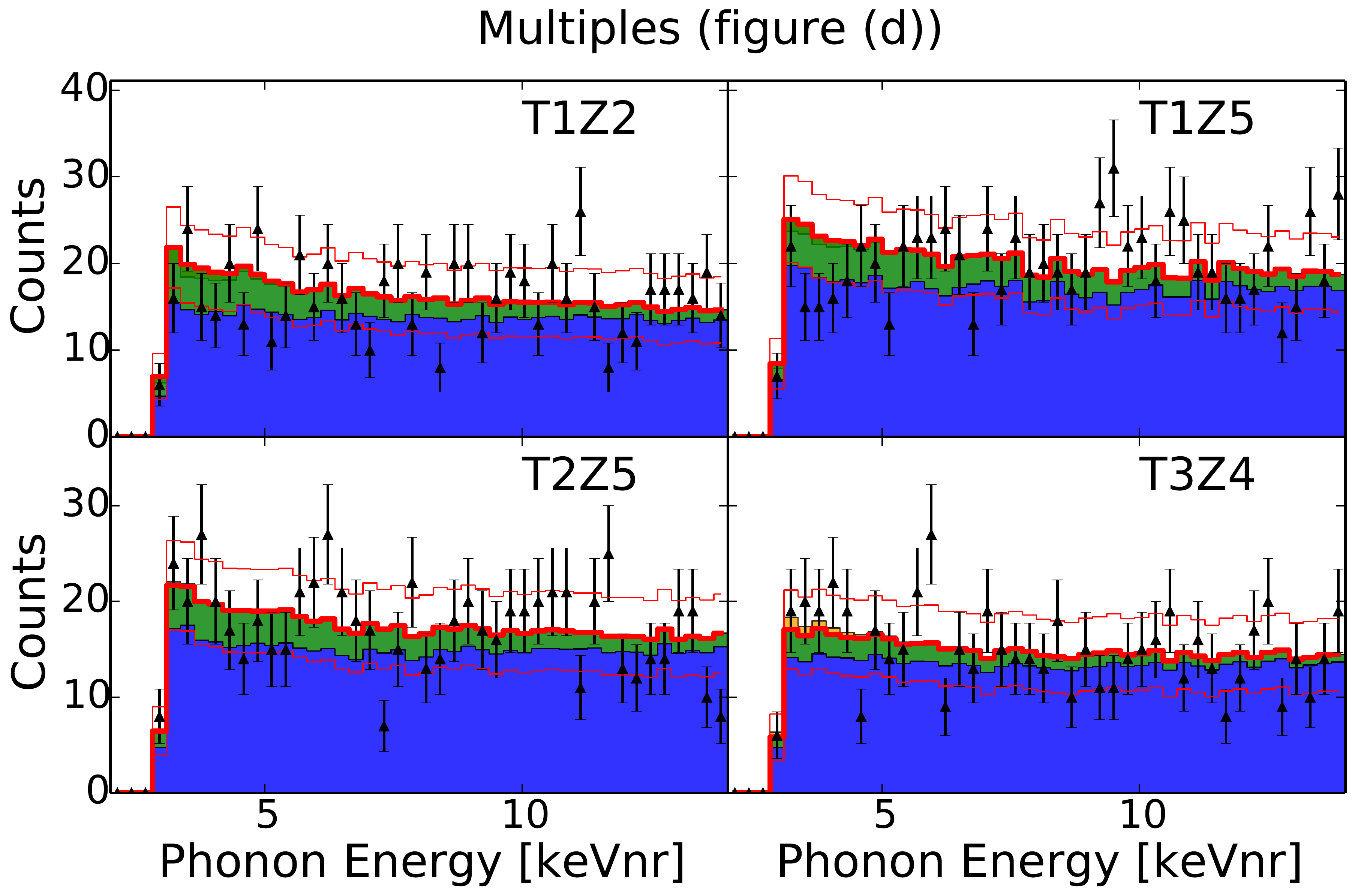}

\caption{(Color online) Stacked histograms of either phonon or ionization energy. 
Figures (a) and (c)  show the  
ML best-fit result to WIMP-search data singles, while figures (b) and (d) show the ML
best fit to WIMP-search data
multiples. Figures (a) and (b) show the projection in ionization energy while (c) and
(d) show the projection in phonon energy. The four canvases in each figure show
the result for each of the four detectors. 
The combined components of the surface-event background model are represented 
by the solid green histogram (legend title:  $^{210}$Pb PDF), while the 
gamma-background model is shown in blue (legend title:  Ba PDF).
The combined probability density functions from
simulation and calibration data are shown as the thick 
line on top of the solid histograms (the thin lines 
indicate the statistical uncertainties), while the
WIMP-search data is shown in black error bars. The orange histogram 
(legend title: NR PDF) represents the
best-fit nuclear recoil-like component. The agreement is good, with T3Z4 having 
the worst fit of the four detectors caused by two histogram bins after the low-energy peak 
(figures (a) and (b)) which are not fit well. 
The low-energy peak of that detector in the multiples is also not fit well, 
which may be a result of mis-calibration due to a lack of penetration-depth 
calibration data for this detector.}
\label{bestFit}%
\end{figure*}%

\begin{figure}
\centering
\includegraphics[width=0.49\textwidth]{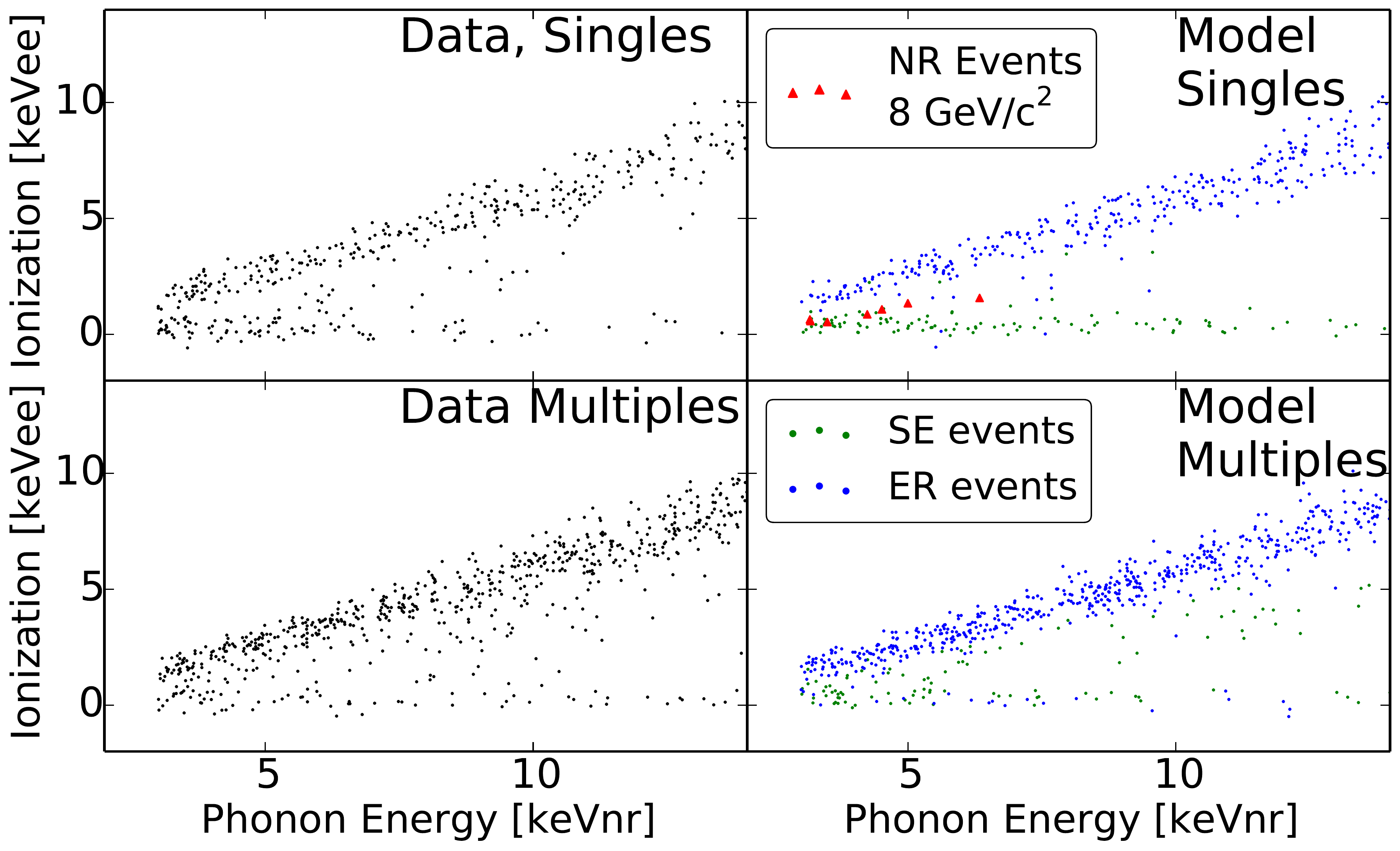}

\caption{(Color online) A comparison between WIMP-search data (left column) 
and the background model (right column) for one detector (T1Z2), with a 
nuclear recoil (NR) component representative of an 8 GeV/c$^2$ WIMP shown 
with the background-model singles. 
The number of points displayed for the background model 
has been reduced to match the number in the WIMP-search data.
The likelihood 
fit is performed in this two-dimensional space of ionization versus~phonon energy.} 
\label{bestFit_scatter}%
\end{figure}%
The gamma background is straightforwardly modeled with Ba calibration events.  
The surface-event background is more difficult due to uncertainty in the locations 
of the radiocontaminants, which are only partially constrained by the observed $^{210}$Po 
alpha decays.
Figure~\ref{AlphaEventPlot} shows energy histograms of the observed alpha 
events that produced most of their ionization signal in the inner sensor 
(detector face) and on the outer sensor (housing).
Unfortunately, uncertainties on the observed alpha rates are large because saturation effects make it 
difficult to reliably reconstruct events at such high energies, and 
because CDMS~II detectors, in contrast to SuperCDMS~iZIPs~\cite{iZipRejection}, cannot 
reliably determine whether an event occurred on the top or the bottom of a detector.
We construct the surface-event component of the detector-face background model by
assuming that 1/4 of the contamination is on each of the four flat surfaces: the
detector's top and bottom and the facing sides of the adjacent detectors.
Consequently, the number of simulated primaries from each detector face is equal.
The number of events expected from the detector housings can be constrained by
counting the number of alpha events that are identified as events occurring on
the outer wall of a detector. Hence the 5 components discussed at the end of
Section~\ref{SE} and in Fig.~\ref{SimContribution} are reduced to a ``Housing"
and a ``Detector Face" component. In the ML fit the relative normalization of 
the different components is fixed accordingly.

%% file: MLH.tex
Based on the surface-event simulations and gamma background estimates 
obtained from Ba calibrations, we perform a maximum likelihood analysis 
to extract the number of WIMP-like events in the WIMP-search data.  
We assume each event in the sample is one of three species: electron 
recoil (ER), surface event (SE), or nuclear recoil (NR). 
The probability density function (PDF) for each species  as a 
function of phonon energy $p$ and ionization energy $q$  
is denoted as ${\cal P}_{XX}(p,q)$. 
${\cal P}_{SE}(p,q)$ is constructed from the surface-event 
simulation to be further discussed in Section~\ref{systematics}, ${\cal P}_{ER}(p,q)$ is constructed from Ba calibration data, and ${\cal P}_{NR}(p,q)$ 
is constructed from a simulated WIMP component at a specific WIMP mass with the ionization yield of 
each event determined from Lindhard theory. 
 All PDFs are 2-dimensional 
binned histograms. 
The NR PDFs are included for both singles and multiples data; since the 
multiples data cannot include a real WIMP signal, including the NR PDF in 
the multiples data provides a useful check for a systematic mischaracterization 
of a background that might result in an apparent WIMP signal in the singles. 
In the fits, the NR component 
is either set identically to zero or allowed to float. 

 The fitting 
is performed using a two-dimensional extended likelihood function 
\begin{equation}
        {\cal L} \;\equiv\;  
        e^{-\bar N}\,
        \prod_{i}^{N} \left(N_{ER} {\cal P}_{ER,i}+N_{SE} {\cal P}_{SE,i}+N_{NR} {\cal P}_{NR,i}\right)~
\label{eqn:lkl}
\end{equation}
where $N$ is the total number of events entering the fit, $N_{XX}$ is 
the fitted number of events per species, and $\bar N\equiv N_{ER}+N_{SE}+N_{NR}$ 
is the total number of fitted events. ${\cal P}_{XX,i}$ is equivalent to ${\cal P}_{XX}(p_i,q_i)$ with $p_i$ and $q_i$ representing the phonon and ionization signal for each event respectively.  The only free parameters in the fits 
are $N_{ER}$, $N_{SE}$, and (when applicable) $N_{NR}$. Note that the 
relative contributions to the SE PDF (shown in Fig.~\ref{SimContribution}) are fixed; only the total number of SE events is allowed to float. 
Furthermore, singles and multiples data are fit independently for each detector. 

The result 
is shown in Fig.~\ref{bestFit}, with the best-fit combined PDF 
shown as the top of the solid histograms for each detector, while low-energy WIMP-search data
are shown as points with error bars. Figure~\ref{bestFit_scatter}
compares WIMP-search data with the background model in the two-dimensional 
space in which fits are performed. The number of simulated events displayed in Fig.~\ref{bestFit_scatter}
has been reduced to be representative of the best-fit value.

%% file: Systematics.tex
The fits done in Section~\ref{MLA} and the background model described in
Section~\ref{bg} depend on parameters that are measured in the experiment itself
in order to properly reproduce the observed data. 
Known systematic effects include possible inaccuracies in the near-surface 
ionization model (see Section~\ref{SE}), inaccuracies in the model of 
the noise and stability characteristics of each detector, and differences 
between the Ba calibration data and the low-energy WIMP-search 
data (see Section~\ref{Gamma_background}).
For this analysis we consider uncertainty in only the near-surface 
ionization model because it has the greatest impact on the background 
model and simultaneously has the most intrinsic uncertainty.

The near-surface ionization model is based on a Monte Carlo study constrained by 
fits to calibration data to determine the ionization yield as a function of distance 
from a detector surface (or ``depth"). This study showed some variation between 
detectors, but also showed that more ionization is collected for events that 
hit the ionization side of 
the detectors compared to the phonon side.
The assumption is that for events that hit the ionization side of
a detector at least 50$\%$ of the electron and hole pairs are collected. The
collection efficiency exponentially increases to 100$\%$ with a 
characteristic depth scale
of $\sim$~1~$\mu{\rm m}$. For events on the phonon side, calibration data are consistent
with the assumption that no charge is collected directly at the surface. 
The amount of charge collected increases exponentially with the
same characteristic depth as the ionization side of the detectors. The last
surface is the detector sidewall. There is currently no calibration data
available for the detector sidewalls. Fitting WIMP-search multiples data we find 
a preference for a characteristic sidewall depth $\sim$5$\times$ smaller than for 
the flat faces. Thus, an event hitting the
detector sidewall will have a higher percentage of its charge collected than
events hitting the flat surfaces.

The reduced charge collection for events near the flat surfaces is
caused by the readout sensors. Phonon 
sensors make up most of the instrumented area and are made from Al
patterned on top of the Ge. Hot charge carriers can back diffuse into 
the Al and therefore be lost.  
Furthermore, charges near one of the flat surfaces will be
attracted to opposite-sign image charges in the Al of the phonon sensors. 
For events on the sidewall, the main mechanism leading to a reduced
charge signal is charge trapping. 
The polishing regime for the sidewalls was not as rigorous as for the flat surfaces.  
Consequently, surface roughness may cause relatively more charge trapping on the sidewalls 
than on the flat surfaces.
In addition, electric field
lines can terminate on the sidewall causing charges not to produce a
signal in the sensor. These effects are partially compensated by the
fact that a free charge near a bare Ge surface experiences a
repulsive force from an image charge of the same sign.

\begin{figure}
\centering
\includegraphics[width=0.49\textwidth]{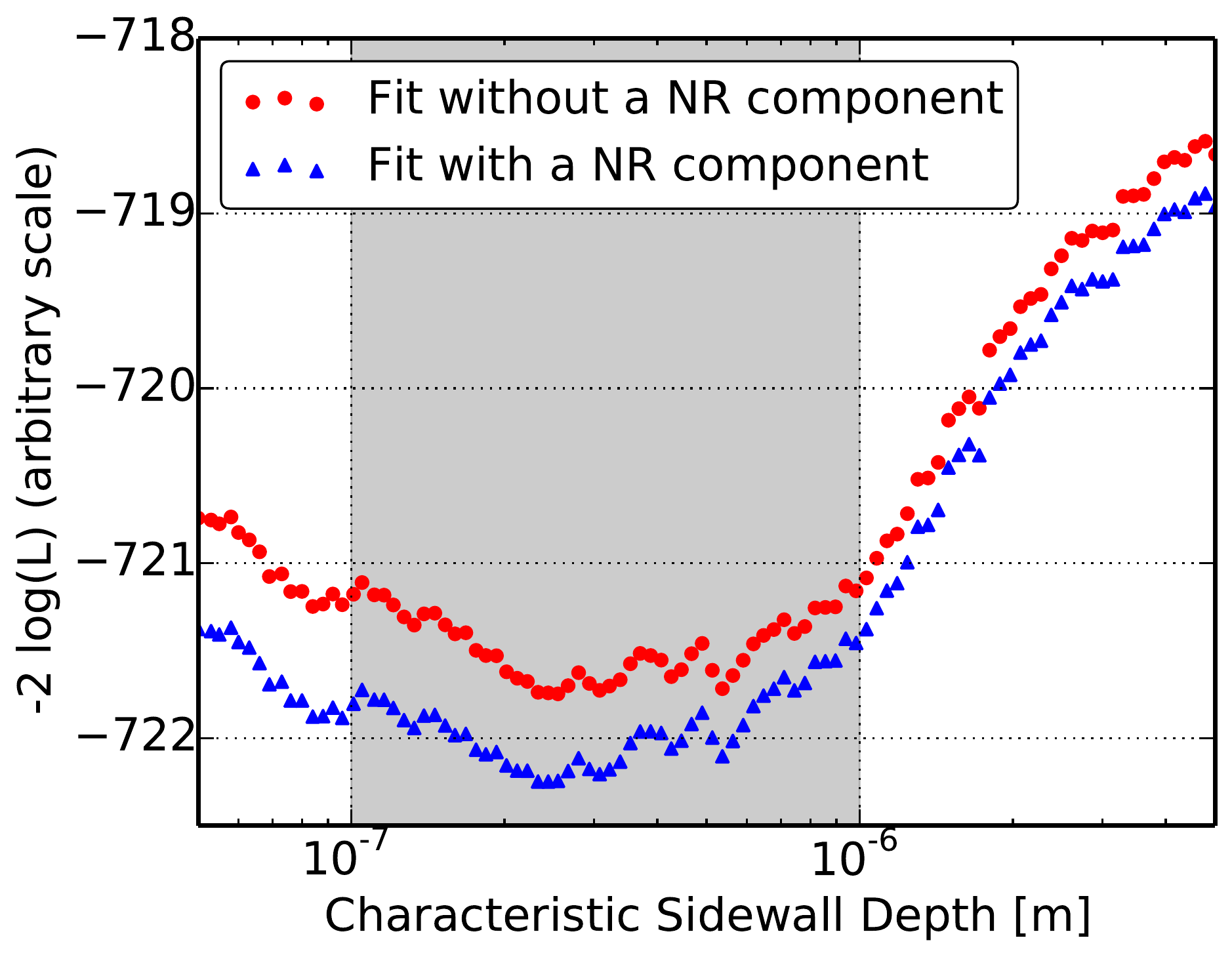}

\caption{(Color online) The best-fit negative log-likelihood value 
for multiples data at various characteristic sidewall depths. A more negative value 
indicates a better fit, although the absolute scale is arbitrary. We show the 
best-fit negative log-likelihood value for both a fit with and without a nuclear recoil 
component. The gray shaded region is the region over which we marginalize 
(using a flat prior) in order to take into account this systematic in the final result.}
\label{systematic}%
\end{figure}

The observed rate on the detector sidewalls is larger than on
the flat surfaces because of higher contamination of the copper surfaces, so the most
sensitive parameter in the model is the characteristic penetration 
depth into the sidewall. We
focus on this parameter in our systematic error evaluation, as an exhaustive
search of the full parameter space is computationally prohibitive. The 
resulting calculation therefore provides a minimum estimate of the systematic 
uncertainty. In Fig.~\ref{systematic} we show the
best-fit likelihood ratio as a function of the assumed sidewall characteristic
depth. We estimate the sidewall depth parameter from fitting multiples data,
which should have no WIMP component, to the background model without a NR component.
Figure~\ref{systematic} indicates that the model is optimized near $\sim$~0.5~$\mu{\rm m}$ 
and becomes worse as the characteristic penetration depth is increased or decreased.
Other parameters of the near-surface ionization model also have an impact on the fit
quality, however much less so.
Performing likelihood ratio tests to compare the fit qualities between  
different characteristic depths for the flat detector faces shows that comparing the 
worst fit to the best fit has a negative log-likelihood value difference that 
is $\sim$5$\times$ smaller than doing the equivalent 
comparison varying the sidewall characteristic depth parameter.
Since the main focus of this work is the ML fitting technique, used by the 
CDMS collaboration for the first time, we have not done an exhaustive 
systematic study, nor
have we attempted to minimize the systematic uncertainties. In future analyses
of SuperCDMS data, we plan to further reduce the systematic effects by using a
detector Monte Carlo simulation~\cite{Geant4_sim} to more accurately simulate detector
physics and thus better determine the detailed detector response. 
Moreover, for our new iZIP detectors, additional calibration data has
already been analyzed~\cite{iZipRejection}, and we plan to perform experimental
studies of sidewall contamination. For the analysis here, we estimate the systematic effects by varying the
sidewall depth parameter between 0.1 and 1~$\mu$m (the shaded region in
Fig.~\ref{systematic}). 

%% file: Results.tex
Having discussed the generation of a background model based on physical knowledge
of our detectors, we are now ready to compute an
exclusion limit on the WIMP-nucleon cross section using the ML technique. 
Systematic effects need to be taken into
account, and as discussed in Section~\ref{systematics} we assume that 
the uncertainty in the
sidewall characteristic depth between 0.1--1\,$\mu$m is a good approximation 
for the systematic effects
encountered in this analysis. 

To calculate a limit on the WIMP cross section we compared our results 
with the results of Monte Carlo simulations that include known numbers of 
WIMP scatters among the background events. These simulations generated 
events from background and WIMPs according to the ${\cal P}_{XX}$ and 
$N_{XX}$ also used in defining the likelihood of Eq.~\ref{eqn:lkl}.  Before producing 
simulations, for each sidewall characteristic depth the maximum likelihood 
value and uncertainty of $N_{ER}$ (and of $N_{SE}$) was found from the 
data with $N_{NR}$ constrained to be zero.  Then for each simulation a 
random value of $N_{ER}$ ($N_{SE}$) was generated according to the maximum 
likelihood and uncertainty.  This random value was in turn taken as the 
expectation for a Poisson random choice of $N_{ER}$ ($N_{SE}$) chosen for 
the simulation.  The simulation's $N_{ER}$ and $N_{SE}$ numbers of events 
were generated for the given sidewall characteristic depth according to 
the already described PDFs, ${\cal P}_{XX}(p,q)$.  Singles data and 
singles PDFs were used for generating MC background of singles, and 
multiples data and multiples PDFs were used for generating MC background 
of multiples.  The expectation value of $N_{NR}$, the number of WIMPs, 
depends on the Galactic halo model, the WIMP mass and cross section, and 
the experiment's detection efficiency, run time, and detector mass, all of 
which were taken from what CDMS assumed and measured in past analyses~\cite{CDMSII_Soudan,CDMSII}.  
Given the expectation of $N_{NR}$, a Poisson random number of 
WIMPs was chosen, and that number of WIMPs was generated according to the 
WIMP PDF.  20000 simulations of background plus WIMPs were produced and 
fit for each sidewall characteristic depth, WIMP mass, and $N_{NR}$ 
expectation value.

\begin{figure*}
\centering
\includegraphics[width=0.49\textwidth]{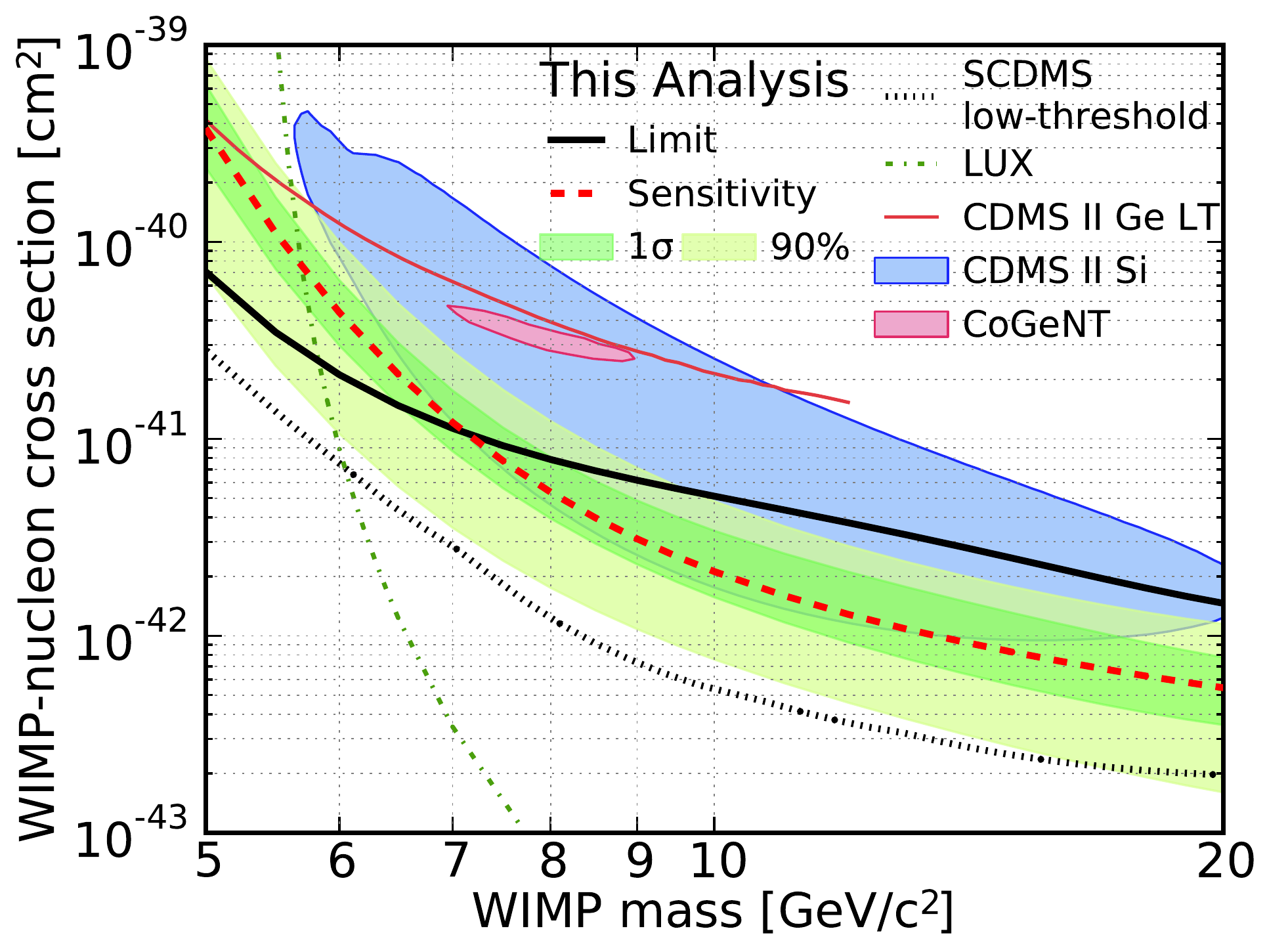}
\includegraphics[width=0.49\textwidth]{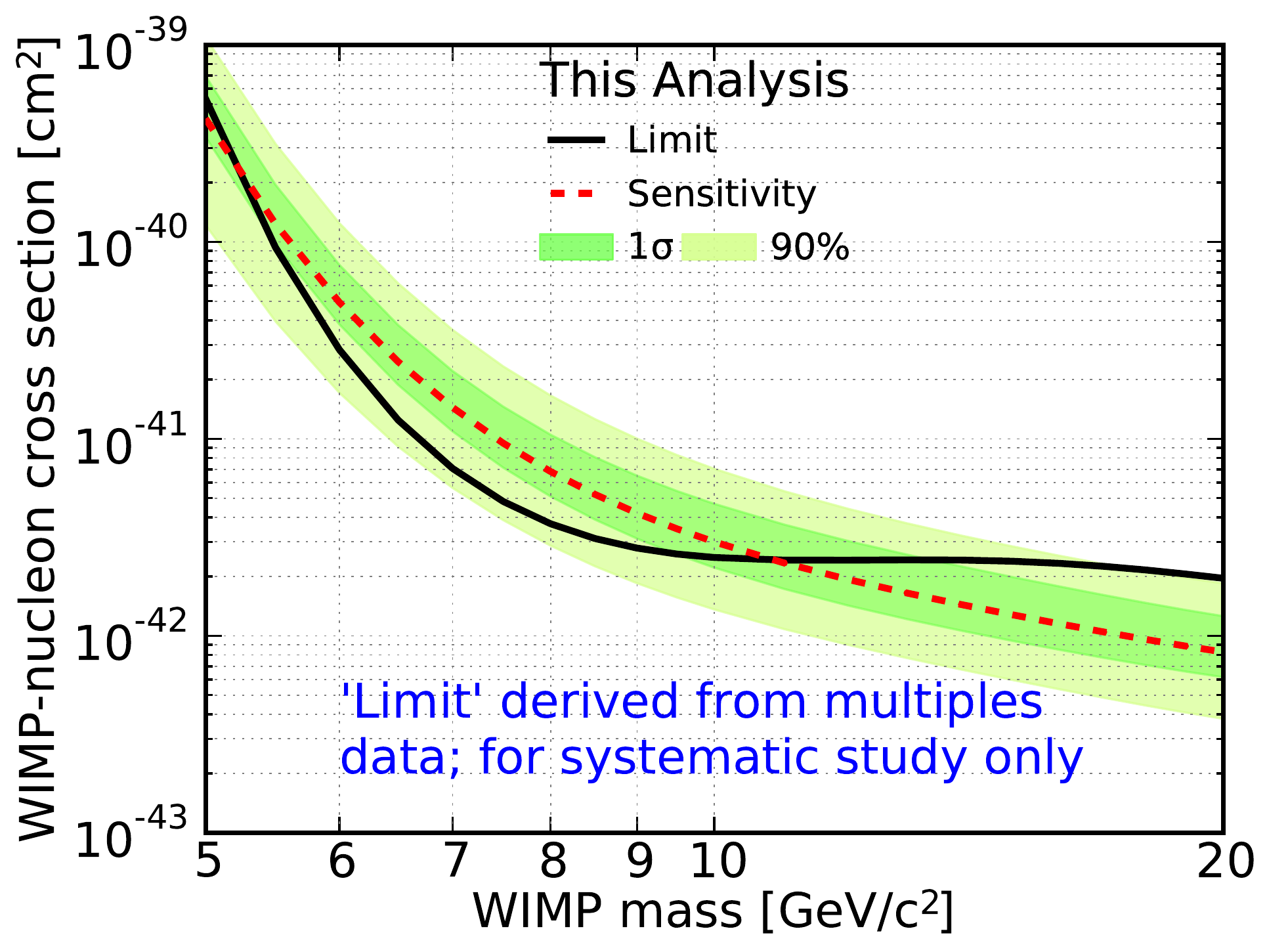}
\caption{(Color online) \textbf{Left panel:} The limit (with standard halo 
assumptions and standard nuclear form factors as 
used in~\cite{CDMSII,SuperCDMS_LT,LUX}) computed for this
analysis is shown as the thick solid black line. 
The thick, dashed (dark-red) line is our best estimate for the expected sensitivity 
of this analysis, with the green (light-green) shaded region directly around it indicating the 
1$\sigma$ (90$\%$) uncertainty. The limit is stronger than
the estimated sensitivity below $\sim$~7~GeV/c$^2$, while at larger WIMP masses
the limit is systematically above the sensitivity, indicating a systematic effect
not yet taken into account. \textbf{Right panel:} This figure shows the limit 
calculated using multiples data instead of singles data. Multiples data do not 
contain WIMPs, and therefore the expected sensitivity and the limit should 
be identical to within statistical fluctuations. 
A similar trend of a stronger than expected limit at lower
WIMP masses and a weaker than expected limit at higher masses is observed, 
indicating that the same systematic effect that is present in the singles is also
present in the multiples data, although to a lesser extent. }
\label{limit_sing}%
\end{figure*}

For each WIMP mass between 5 and 20 GeV/c$^2$, and each sidewall 
characteristic depth, we began the upper limit calculation by first 
finding the maximum likelihood best-fit number of WIMPs in the singles 
data, $N_{NR}$.  For various values of the WIMP cross section we performed 
20000 Monte Carlo (MC) simulations of the experiment, each with an assumed 
WIMP cross section, and found $N_{exceed}$, the number of simulations for 
which the maximum likelihood best-fit $N_{NR}$ exceeded the value found in 
the singles data.  In Fig.~\ref{limit_sing} the black curve indicating the 90\% upper 
limit for each WIMP mass shows the cross section for which 90\% of MC 
simulations found at least as many events as were found in the real data. 
In order to include a crude estimate of the effect of systematics, we 
marginalized over sidewall characteristic depth.  The calculation of 
$N_{exceed}$ was done for each sidewall characteristic depth, and the 
resulting values were summed over 50 uniformly spaced sidewall 
characteristic depths from 0.1--1.0\,$\mu$m.  The 90\% upper limit for 
the WIMP mass under consideration was then taken to be the cross section 
for which this total sum of the $N_{exceed}$ values was at least 90\% of 
the total number of MC simulations over all characteristic sidewall 
depths.  This procedure weakens the limit, because for large sidewall 
depths ($\sim 1\ \mu$m) the ML fit number of WIMPs from the data increases 
significantly and dominates the limit.

A 90\% sensitivity curve was also computed.  To obtain the dashed (red) 
curve in Fig.~\ref{limit_sing} a calculation was done similar to that for the 90\% upper 
limit, but with the singles data value of $N_{NR}$ replaced by values 
found from fits to MC simulations generated without WIMPs.  Since the MC 
fit values of $N_{NR}$ vary from one simulation to another, the 90\% upper 
limits vary.  This variation is indicated by 1$\sigma$ and 90\% regions 
about the sensitivity curve (darker and lighter green bands).

In order to test our methods on data that we know is free of WIMPs, the 
upper limit and sensitivity results were also calculated from multiples 
data treated as if WIMPs could be present.
The sensitivity and limit are shown in Fig.~\ref{limit_sing}.

The limit at low WIMP masses is
stronger than the expected sensitivity, while at high WIMP masses it is weaker.
The fact that the limit is above the 90$\%$ sensitivity band (the light green band
in Fig.~\ref{limit_sing}) points to either a possible WIMP signal (if limits set
by other experiments are not taken into consideration) or more likely a
deficiency in the background model. The WIMP significance above 10~GeV/c$^2$ is
$\sim$~2$\sigma$. In order to check the background model we can also produce a
limit plot using multiples data instead of singles data. Of course multiples
data do not contain any WIMP signal, and therefore the sensitivity should
agree with the limit within statistical fluctuations for a perfect background
model. This is shown in Fig.~\ref{limit_sing} in the right panel. While there
does not appear to be a fluctuation to a stronger than expected limit at low
WIMP masses, the trend seen in the singles limit of a weaker limit at higher
WIMP masses is repeated in the multiples data. This
result suggests that small residual systematics in the background model are
responsible for weakening the limit for higher WIMP masses.
\newline

The power of performing a likelihood analysis is illustrated in Fig.~\ref{limit_sing} 
by comparing the ``CDMS II LT" curve, based on analysis of the 
same data without background subtraction~\cite{CDMSII}, to the limit of this analysis. 
We see a factor of $\sim$5 improvement. 
Another check of the power of a likelihood analysis is to compare the sensitivity of this
analysis to the sensitivity of the SuperCDMS low-threshold result~\cite{SCDMS_LT}.
Both analyses are background-limited, but the background in the CDMS~II data analyzed 
here is considerably higher than the background in the SuperCDMS data.
However, with the advanced analysis technique presented here we reach a sensitivity that 
almost scales with the exposure (for a 4.5 times larger exposure, SuperCDMS increases the 
sensitivity by a factor of 5). This result suggests that
the technique presented here may help to improve current SuperCDMS limits, as
well as those of future experiments.

%% file: ComparisonToCF.tex
\begin{table*}[!htp]
\begin{tabular}{|c||cc||cc|cc||cc|cc||cc|}
\hline
          & \multicolumn{2}{c||}{Collar \& Fields~\cite{CF}} & \multicolumn{10}{c|}{CDMS Collaboration Analyses}\\
          \hline
         & \multicolumn{2}{c||}{Singles} & \multicolumn{2}{c|}{Singles Data}&\multicolumn{2}{c||}{Multiples Data}&\multicolumn{2}{c|}{Singles Simulation}&\multicolumn{2}{c||}{Multiples Simulation}&\multicolumn{2}{c|}{Likelihood}\\
Detector &    $N_{NR}$   &  $A_{2}$  & $N_{NR}$   &  $A_{2}$   & $N_{NR}$   &  $A_{2}$ & $N_{NR}$   &  $A_{2}$   & $N_{NR}$   &  $A_{2}$ & Singles & Multiples\\
\hline
T1Z2    &   $33\pm 9$  &  $0.6\pm0.1$  & $23\pm 8$  &  $0.4\pm0.1$& $ 10\pm 6 $  &  $1.5\pm0.7$ &$22\pm 9$ &  $0.7\pm 0.3$ & $7\pm 10$ &  $0.9\pm 0.7$ &$8\pm13$&$-7\pm10$ \\
T1Z5    &   $18\pm 6$  &  $0.7\pm0.3$  & $16\pm 6$  &  $0.5\pm0.2$& $17\pm 8$  &  $0.2\pm0.2$ & $13\pm8$ & $0.7\pm0.4$ & $11\pm10$ &  $0.9\pm 0.6$& $-1\pm11$&$-7\pm11$ \\
T2Z3    &   $37\pm 19$  &  $0.7\pm0.2$ & $30\pm18$  &  $0.9\pm0.4$& $45\pm13$  &  $0.5\pm0.2$ &&&&&&\\
T2Z5    &   $26\pm 14$  &  $0.8\pm0.4$ & $30\pm13$  &  $0.9\pm0.4$& $83\pm16$  &  $0.4\pm0.1$ &  $30\pm9$ & $0.7\pm0.3$ & $32\pm11$ &  $0.8\pm 0.3$&$6\pm18$ & $2\pm20$\\
T3Z2    &   $26\pm 10$  &  $0.7\pm0.2$  & $14\pm12$  &  $1.1\pm0.7$& $17\pm12$  &  $0.5\pm0.4$&&&&&&\\
T3Z4    &   $12\pm 4$  &  $0.6\pm0.2$  & $10 \pm 4$  &  $0.6\pm0.2$& $8 \pm 5$  &  $0.6\pm0.4$ & $5\pm5$ &  $0.9\pm0.5$& $6\pm7$ &  $0.9\pm 0.6$&$9\pm13$&$6\pm13$ \\
T3Z5    &   $4\pm 10$  &  $2.0\pm2.4$   & $9 \pm10$  &  $1.8\pm1.2$& $91\pm18$  &  $0.4\pm0.1$&&&&&&\\
T3Z6    &   $22\pm 11$  &  $0.7\pm0.4$ & $24\pm 8$  &  $0.6\pm0.3$& $ 2\pm 3$  &  $0.0\pm0.1$&&&&&&\\
\hline
Best Four &   $89\pm 18$  &   & $79\pm17$  &  & $ 118\pm20$  &  &$72\pm13$&$0.6\pm0.1$&$68\pm16$&$0.7\pm0.2$ &$22\pm28$&$-6\pm29$\\
All Dets  &   $178\pm 32$  &  $0.7\pm0.1$ & $153\pm33$  &  $0.6\pm0.1$& $ 231\pm34$  &  $0.6\pm0.1$&$72\pm13$&$0.6\pm0.1$&$68\pm16$&$0.7\pm0.2$ &$22\pm28$&$-6\pm29$\\
\hline
\end{tabular}
\caption{\label{tab:CFComparison} The number of NR-like events and the NR
exponential constant extracted from the WIMP-search data using the analytic
fit (described in the text).  We also provide the equivalent numbers from~\cite{CF}. 
The ``Simulation" columns show what happens if we fit our background model 
using the Collar-Fields PDFs defined in Equation~\ref{eqn:cf}. The observed excess in this case 
is on par with the 
observed excess in WIMP-search data. The last two columns (labeled ``Likelihood") 
show the number of WIMP-like 
events preferred (for a mass of 8~GeV/c$^2$) when WIMP-search data is fit with our
background model using a sidewall depth of 0.3~$\mu{\rm m}$ (the optimal value, 
see Fig.~\ref{systematic}).
We chose a mass of 8~GeV/c$^2$ because that is the preferred value for the Collar-Fields 
type analysis, as well as other closed contours (see Fig.~\ref{limit_sing}). Using our 
background model we do not observe an excess at this WIMP mass (see last 2 columns).}
\end{table*}

For comparison, we also perform a maximum likelihood fit to the WIMP-search data
using analytic PDFs similar to those used by Collar and Fields~\cite{CF}.  The
form of the likelihood function is similar to  Equation~(\ref{eqn:lkl}) except the three components are ER, ZC, and NR explicitly written as:
\begin{equation}
        {\cal L} \;\equiv\;  
        e^{-\bar N}\,
        \prod_{i}^{N} \left(N_{ER} {\cal P}_{ER,i}+N_{ZC} {\cal P}_{ZC,i}+N_{NR} {\cal P}_{NR,i}\right). 
\label{eqn:lklcf}
\end{equation}
Instead of using histograms the PDFs are two-dimensional functions in $(q,p)$. 
Specifically, the PDFs are of the form:
\begin{equation}
{\cal P}_{XX}(q,p)=\exp\left(-A_{2,XX} p\right) \exp\left(\frac{-\left(q-C_{XX}\left(p\right)\right)^2}{S_1^2+S_{2}C_{XX}\left(p\right)}\right),
\label{eqn:cf}
\end{equation}
where $C_{XX}$ is a polynomial describing the mean $q$ of the recoil band as a function of $p$. 
We use a polynomial of order 0 (1,2) for the ZC (ER, NR) band, respectively. 
For the fit to the data from individual detectors, the coefficients $C_{XX}$ are 
fixed to values obtained from calibration samples from that detector.  
We also perform fits to the singles and multiples samples where we combine data 
from all of the detectors; in these fits the $C_{XX}$ are allowed to float.  Following
\cite{CF}, the ER PDF is slightly modified from Equation~(\ref{eqn:cf})
to include surface events with incomplete charge collection (the
so-called ``Crystal Ball" function~\cite{crystalball}). Note that this treatment is 
different from the model described in Section~\ref{FullBackgroundModel} where the events with
 zero and incomplete charge collection are included in the same PDF.  
The results of the fits using the analytic PDFs are summarized in Table~\ref{tab:CFComparison}. They 
are in reasonable agreement with Ref.~\cite{CF} considering there are some
differences in the datasets used.  In particular, we observe a significant improvement 
(4.4$\sigma$) to the fit to singles data when a NR
component is included.  However, we also see a significant 
improvement (5.2$\sigma$) when we perform these fits to the multiples data (also shown in
Table~\ref{tab:CFComparison}). 

In addition, we performed fits using the
analytic model to an ensemble of toy MC datasets generated from the best fit of our background model (without a NR component) to the WIMP-search data (as detailed in Section  \ref{results}). 
The average fitted number of NR events and $A_{2,NR}$ 
(from 100 datasets) are shown in the columns of Table~\ref{tab:CFComparison},
labeled ``Singles Simulation" and ``Multiples Simulation."  We see good agreement
when comparing the fitted WIMP parameters between these toy (WIMP-free) datasets
and the fits to data. These two factors, significant WIMP components in the 
multiples data and in toy datasets generated from our physics-based model (without a NR component), lead us to conclude that the excess NR-like events identified by the Collar-Fields analytical model are not true nuclear recoils but are instead due to an inability of this parametrization to adequately describe the background. Finally, the Table~\ref{tab:CFComparison} columns
labeled ``Likelihood" show results of fits to WIMP-search data using our background model
plus an 8~GeV/c$^{2}$ WIMP component. 
There are $< 1\sigma$ WIMP-like upward (downward) fluctuations  in singles (multiples) data. 
It is clear that our background model performs significantly better than the \textit{ad hoc} parameterization from~\cite{CF}. We believe the superiority of our background model can be attributed to the inability of the  \textit{ad hoc} functions to 
properly describe the surface-event background from the $^{210}$Pb decay chain.  

%% file: Conclusion.tex
We presented the results of a Maximum Likelihood fit to the low-energy  CDMS~II Ge WIMP-search
data.  We used a background model derived from detector simulations and
calibrations from the known contributing sources.  We observe no significant 
NR component in our data and set a limit on the WIMP-nucleon cross section as a function of WIMP mass that is
generally 5$\times$ stronger than our previous analysis of these data, 
which did not include any background subtraction~\cite{CDMSII}.  This result demonstrates the power of the
ML technique for low-threshold WIMP searches. We also performed a fit to the
dataset using the {\it ad hoc} analytic PDFs suggested by Collar and Fields~\cite{CF}, that produces
a significant excess of NR-like events in this dataset.  Using their method, we 
reproduce their results for the single-scatter data but also observe a significant
excess in multiple-scatter data, leading us to conclude that their analytical model
is insufficient to describe the backgrounds.

%% file: Acknowledgements.tex
The CDMS collaboration gratefully acknowledges the contributions of numerous engineers and technicians; 
we would like to especially thank Dennis Seitz, Jim Beaty, Bruce Hines, Larry Novak, Richard Schmitt, 
Astrid Tomada, and John Emes. In addition, we gratefully acknowledge assistance from the staff of 
the Soudan Underground Laboratory and the Minnesota Department of Natural Resources. This work is 
supported in part by the National Science Foundation, by the United States Department of Energy, by 
NSERC Canada, and by MultiDark (Spanish MINECO). Fermilab is operated by the Fermi Research Alliance, 
LLC under Contract No. De- AC02-07CH11359. SLAC is operated under Contract No. DE-AC02-76SF00515 with 
the United States Department of Energy.

%% file: CDMSII_LLH_Paper.bbl
\begin{thebibliography}{10}

\bibitem{Planck}
{Planck Collaboration}, P.~A.~R. {Ade}, and et~al., ``{Planck 2013 results. I.
  Overview of products and scientific results},'' {\em arXiv:1303.5062}, 2013.

\bibitem{WIMPsRef}
G.~Steigman and M.~Turner, ``Cosmological constraints on the properties of
  weakly interacting massive particles,'' {\em Nucl. Phys. B}, vol.~253,
  pp.~375--386, 1985.

\bibitem{CDMSII_Soudan}
D.~S. Akerib {\em et~al.}, ``{Low-threshold analysis of CDMS shallow-site
  data},'' {\em \prd}, vol.~82, no.~12, p.~122004, 2010.

\bibitem{CDMSII}
Z.~Ahmed {\em et~al.}, ``Results from a low-energy {Analysis} of the {CDMS II}
  {Germanium} {Data},'' {\em Phys. Rev. Lett.}, vol.~106, p.~131302, 2011.

\bibitem{DAMA}
R.~Bernabei {\em et~al.}, ``{First results from DAMA/LIBRA and the combined
  results with DAMA/NaI},'' {\em Eur. Phys. J.}, vol.~C56, pp.~333--355, 2008.

\bibitem{COGENT}
C.~E. Aalseth {\em et~al.}, ``Search for an annual modulation in a p-type point
  contact germanium dark matter detector,'' {\em Phys. Rev. Lett.}, vol.~107,
  p.~141301, 2011.

\bibitem{CDMS_SI}
R.~Agnese {\em et~al.}, ``{Silicon Detector Dark Matter Results from the Final
  Exposure of CDMS II},'' {\em Phys. Rev. Lett.}, vol.~111, p.~251301, 2013.

\bibitem{modulation}
Z.~Ahmed {\em et~al.}, ``{Search for annual modulation in low-energy CDMS~II
  data},'' {\em arXiv:1203.1309}, 2012.

\bibitem{SuperCDMS_LT}
R.~Agnese {\em et~al.}, ``{CDMSlite: A Search for Low-Mass WIMPs using
  Voltage-Assisted Calorimetric Ionization Detection in the SuperCDMS
  Experiment},'' {\em Phys. Rev. Lett.}, vol.~112, p.~041302, 2014.

\bibitem{LUX}
D.~Akerib {\em et~al.}, ``{First results from the LUX dark matter experiment at
  the Sanford Underground Research Facility},'' {\em Phys. Rev. Lett.},
  vol.~112, p.~091303, 2014.

\bibitem{CF}
J.~I. {Collar} and N.~E. {Fields}, ``{A Maximum Likelihood Analysis of
  Low-Energy CDMS Data},'' {\em arXiv:1204.3559}, 2012.

\bibitem{Luke}
P.~N. Luke, ``{Voltage assisted calorimetric ionization detector},'' {\em Appl.
  Phys. Lett.}, vol.~64, p.~6858, 1988.

\bibitem{KS}
M.~A. Stephens, ``Edf statistics for goodness of fit and some comparisons,''
  {\em J. Am. Statist. Assoc.}, vol.~69, no.~347, pp.~730--737, 1974.

\bibitem{Pb206_Redl}
P.~Redl, ``Accurate {Simulations of} {$^{206}$Pb} {Recoils} in {SuperCDMS},''
  {\em Journal of Low Temperature Physics}, pp.~1--6, 2014.

\bibitem{Geant-B}
J.~Allison {\em et~al.} {\em IEEE Trans. Nucl. Sci.}, vol.~53, pp.~270--278,
  2006.

\bibitem{iZipRejection}
R.~Agnese {\em et~al.}, ``{Demonstration of Surface Electron Rejection with
  Interleaved Germanium Detectors for Dark Matter Searches},'' {\em Appl. Phys.
  Lett.}, vol.~103, p.~164105, 2013.

\bibitem{SNR}
M.~H. {Mendenhall} and R.~A. {Weller}, ``{An algorithm for computing screened
  Coulomb scattering in G EANT4},'' {\em Nucl. Instrum. Methods Phys. Res. B},
  vol.~227, pp.~420--430, 2005.

\bibitem{SRIM}
J.~F. {Ziegler}, M.~D. {Ziegler}, and J.~P. {Biersack}, ``{SRIM - The stopping
  and range of ions in matter (2010)},'' {\em Nucl. Instrum. Methods Phys. Res.
  B}, vol.~268, pp.~1818--1823, 2010.

\bibitem{Geant4_sim}
D.~Brandt, R.~Agnese, P.~Redl, K.~Schneck, M.~Asai, M.~Kelsey, D.~Faiez,
  E.~Bagli, B.~Cabrera, R.~Partridge, T.~Saab, and B.~Sadoulet,
  ``{Semiconductor phonon and charge transport Monte Carlo simulation using
  Geant4},'' {\em arXiv1403.4984}, 2014.

\bibitem{eEV_cite}
X.-F. {Navick}, M.~{Chapellier}, F.~{D{\'e}liot}, S.~{Herv{\'e}}, and
  L.~{Miramonti}, ``{320 g ionization-heat bolometers design for the EDELWEISS
  experiment},'' {\em Nucl. Instrum. Methods Phys. Res. A}, vol.~444,
  pp.~361--363, Apr. 2000.

\bibitem{Lindhard}
P.~Smith and J.~Lewin, ``{Dark Matter Detection},'' {\em Phys. Rept.},
  vol.~187, p.~203, 1990.

\bibitem{DeadLayer}
V.~Mandic {\em et~al.}, ``Study of the dead layer in germanium for the {CDMS}
  detectors,'' {\em Nucl. Instrum. Methods Phys. Res. A}, vol.~520, no.~1–3,
  pp.~171 -- 174, 2004.
\newblock Proceedings of the 10th International Workshop on Low Temperature
  Detectors.

\bibitem{SCDMS_LT}
R.~Agnese {\em et~al.}, ``{Search for Low-Mass WIMPs with SuperCDMS},'' {\em
  Phys. Rev. Lett.}, vol.~112, p.~241302, 2014.

\bibitem{crystalball}
M.~Oreglia, ``{A Study of the Reactions $\psi^\prime \to \gamma \gamma
  \psi$},'' {\em Ph.D, thesis, SLAC, SLAC-R-0236, Appendix D}, 1980.

\end{thebibliography}
